%% file: sumrules.tex
\documentclass[11pt,paper]{JHEP3}
\usepackage{epsfig}
\usepackage{latexsym}
\usepackage{amssymb}
\usepackage{booktabs}
\newcommand{\be}{\begin{equation}}
\newcommand{\ee}{\end{equation}}
\newcommand{\ba}{\begin{eqnarray}}
\newcommand{\ea}{\end{eqnarray}}
\newcommand{\bi}{\begin{itemize}}
\newcommand{\ei}{\end{itemize}}
\newcommand{\ben}{\begin{enumerate}}
\newcommand{\een}{\end{enumerate}}

\newcommand{\tr}{{\rm Tr\,}}
\newcommand{\re}{\mathop{\rm Re}}

\newcommand{\half}{{\textstyle\frac{1}{2}}}

\newcommand{\<}{\langle}
\renewcommand{\>}{\rangle}
\newcommand{\eq}{Eq.~}

\newcommand{\fig}{Fig.~}

\newcommand{\la}{\label}

\newcommand{\txts}{\textstyle}

\newcommand{\im}{\mathop{\rm Im}}

\newcommand{\Nt}{N_{\tau}}

\renewcommand{\O}{{\cal O}}

\def\tr{{\rm tr}}

\def\eeq{\end{equation}}
\def\half{\frac{1}{2}}

\title{The Bulk Channel in Thermal Gauge Theories}

\author{
Harvey~B.~Meyer \\
CERN Physics Department\\
1211 Geneva 23, Switzerland \\
and\\
Center for Theoretical Physics\\
Massachusetts Institute of Technology\\
Cambridge, MA 02139, U.S.A.\\
E-mail: \email{hmeyer@mail.cern.ch}
}

\keywords{Lattice Gauge Field Theories, Thermal Field Theory}
\preprint{CERN-PH-TH/2010-037\\
          MIT-CTP 4121}

\abstract{
We investigate the thermal correlator of the trace of the 
energy-momentum tensor in the SU(3) Yang-Mills theory.
Our goal is to constrain the spectral function in that channel,
whose low-frequency part determines the bulk viscosity.
We focus on the thermal modification of the spectral function,
$\rho(\omega,T)-\rho(\omega,0)$. 
Using the operator-product expansion
we give the high-frequency behavior of this difference 
in terms of thermodynamic potentials.
We take into account the presence 
of an exact delta function located at the origin,
which had been missed in previous analyses.
We then combine the bulk sum rule and a Monte-Carlo evaluation 
of the Euclidean correlator to determine the intervals of frequency where 
the spectral density is enhanced or depleted by thermal effects.
We find evidence that the thermal spectral density 
is non-zero for frequencies below the scalar glueball mass $m$
and is significantly depleted for $m\lesssim\omega\lesssim 3m$.
}

\begin{document}
\section{Introduction}
Understanding the emergence of bulk properties and collective
behavior from the strong interaction of particle physics
is a goal that has been pursued vigorously in recent years.
Experimentally, heavy-ion collisions have provided a way 
to heat up nuclear matter to temperatures at which quarks and gluons 
are expected to dominate its macroscopic
properties~\cite{Arsene:2004fa,Back:2004je,Adcox:2004mh,Adams:2005dq}.
Theoretically, both analytic~\cite{Kapusta:2006pm}
and numerical methods~\cite{DeTar:2009ef}
are being used to unravel the intricate dynamics of QCD 
at finite temperature and baryon density.

At temperatures of a few hundred MeV and small baryon density, 
lattice QCD remains the tool of choice to determine the 
equilibrium properties of the system.
As we shall discuss in detail,
access to the near-equilibrium properties,
such as the transport properties, is limited 
in this approach, because lattice QCD employs the 
Euclidean formulation of thermal field theory.
Real-time properties can thus only be determined
by analytic continuation, which for numerical 
purposes represents an ill-posed problem.
However, it is not excluded that we 
can learn about the gross features 
of the spectral density by suitably 
combining numerical and analytic methods.
The purpose of this paper is to illustrate how 
such a combination of methods can be put to practice,
taking as an example the bulk channel in the 
SU(3) Yang-Mills theory.

One of the lasting legacies
of the Relativistic Heavy Ion Collider (RHIC) experiments
is the success of the hydrodynamic description of 
the collisions, see for instance the review~\cite{Teaney:2009qa}. 
The early agreement between ideal hydrodynamics
and experiment was gradually refined to incorporate
the dissipative effects of shear viscosity $\eta$,
the  three-dimensional 
geometry~\cite{Romatschke:2007mq,Song:2007ux,Dusling:2007gi}
and to estimate 
the sensitivity to initial conditions~\cite{Luzum:2008cw,Heinz:2009cv}.
Most recently, the effects of bulk viscosity $\zeta$ have been
investigated~\cite{Song:2009rh,Rajagopal:2009yw}.
At temperatures sufficiently high for weak coupling 
methods to be applicable, the bulk viscosity is 
three orders of magnitude smaller than 
the shear viscosity~\cite{Arnold:2006fz}.
However, by analogy with many condensed matter 
systems (see for instance~\cite{Kapusta:2008vb}),
where a maxixum in the bulk viscosity is observed 
near liquid-gas phase transitions, one generally
expects the bulk viscosity to become appreciable, 
if anywhere, near the rapid crossover 
between the confined and the deconfined phases.
It has been suggested that this sudden onset of 
dissipation under expansion could trigger the 
freeze-out in heavy-ion collisions 
due to cavitation~\cite{Rajagopal:2009yw};
the breakup of a fluid occurs when a component of the 
stress-energy tensor, say $T_{zz}$, turns negative.
The behavior of $\zeta$ near the QCD crossover was studied
about two-and-a-half years ago, 
first using a QCD sum rule and a crude model 
for the bulk spectral function~\cite{Kharzeev:2007wb,Karsch:2007jc}, 
and subsequently using 
Euclidean correlators obtained by Monte-Carlo methods~\cite{Meyer:2007dy}.
Both types of analyses missed the fact that a homogeneous 
fluctuation in $T_{\mu\mu}$ has a component which 
takes arbitrarily long to relax to equilibrium.
It is therefore necessary to revisit these 
analyses, and this constitutes the second motivation for this work.
A study based on AdS/CFT methods~\cite{Gubser:2008yx}
also found an increase in the bulk viscosity near $T_c$,
however less pronounced than suggested by 
\cite{Kharzeev:2007wb,Meyer:2007dy}.

In the near future, the low-energy runs at RHIC,
and further ahead, the heavy-ion experiments at FAIR
will explore the QCD diagram at larger net baryon
density. It is widely thought, although not beyond
doubt, that a chiral critical point exists in the 
$T$ vs. $\mu_B$ plane~\cite{Stephanov:2004wx}.
The associated critical exponents~\cite{Son:2004iv}
indicate that the bulk viscosity 
should exhibit a strong divergence 
near the critical point.
If this increase in $\zeta$ is not too closely 
localized around $T_c$, it could be one of the signatures
of the critical point.

The outline of this paper is as follows.
We start in section \ref{sec:review} with a review 
of retarded and Euclidean correlators, establishing
the precise relations between them. Details are 
collected in the appendix. 
In section (\ref{sec:bulk}) we turn to the specific 
case of the bulk channel in the SU($N$) gauge theory.
We describe how to use the bulk sum rule as a 
constraint on the vacuum-subtracted 
spectral density $\Delta\rho$, and present the operator
product expansion (OPE) prediction for the large-frequency
behavior of the spectral density.
In section \ref{sec:num} we give a numerical application
of our results for $N=3$. We first present data deep in the 
confined phase, as well as in the deconfined phase. 
The sum rule is combined with lattice data to yield
information on the changes in the spectral density 
that occur between the confined and the deconfined phases.
We end with the conclusions of this study.

\section{A review of finite-temperature correlators\la{sec:review}}
In this section, we give the definitions of 
the retarded correlator and the Euclidean correlator,
and show their relation in detail.
This allows us to fix our conventions, but also 
to highlight a subtlety that arises in the 
zero-frequency case.
The retarded correlator, evaluated at a purely imaginary 
frequency, is the analytic continuation of the Euclidean 
correlator, which is only available at the Matsubara frequencies.
We show that the odd derivatives of the retarded correlator
along the imaginary axis, 
evaluated at the origin, coincide with the derivatives of the 
spectral density at the origin. The even derivatives,
on the other hand, correspond to integrals over all frequencies 
of the spectral density.
We illustrate the relations that we derive
on the example of the bulk channel in SU($N$) gauge theory.
Working in frequency space rather than in time-coordinate space,
we show the relation between the bulk viscosity and the 
frequency-space Euclidean correlator, \fig(\ref{fig:sketch}).

We work in the canonical ensemble, i.e.
expectation values of operators are defined by
\be
\<\hat\O\> = \frac{1}{Z}\tr\{e^{-\beta \hat H}\, \hat\O\},
\ee
where the partition function $Z(\beta)$ is such that
 $\<1\>=1$ and $\beta=1/T$ is the inverse temperature.
In the following we consider a generic Hermitian operator
 ${\cal O}$.
We use the Heisenberg representation, with 
\be
\hat\O(t) \equiv e^{i\hat H t} \hat\O(0) e^{-i\hat H t}\,.
\ee
We start by introducing the real-time correlators,
\be
G_R(t) = i\theta(t) \<[\hat\O(t), \hat\O(0)]\>\,.
\la{eq:defG_R-main}
\ee
Its Fourier transform 
\be
\tilde G_R(\omega) = i\int_0^\infty dt\, e^{i\omega t} 
\<[\hat\O(t), \hat\O(0)]\>.
\la{eq:defG_Rtilde-main}
\ee
plays an important role in the following.
With these conventions we note that $\tilde G_R$
is analytic in the half complex plane $\im(\omega)>0$.

The Euclidean correlator is defined as
\be
G_E(t) =\< \hat\O(-it)\, \hat\O(0)\>.
\la{eq:GE}
\ee
It obeys the Kubo-Martin-Schwinger relation
\be
G_E(\beta-t) = G_E(t)\,,
\ee
and can be expressed as a Fourier series
on the interval $0\leq t<\beta$,
\ba
G_E(t) &=& T\sum_{\ell\in{\bf Z}} G_E^{(\ell)}\, e^{i\omega_\ell t}\,,
\\
G_E^{(\ell)} &=&  \int_0^\beta  dt e^{-i\omega_\ell t} G_E(t)\,,
\la{eq:GEell}
\ea
where $\omega_\ell= \omega_M\cdot \ell = 2\pi T\,\ell$.

The spectral representation of the correlators is useful 
to prove simple relations between the Euclidean and 
real-time correlators. 
One finds that (see the appendix)
\be
G_R(i\omega_\ell) = G_E^{(\ell)} \,, \quad
\ell\neq 0\,.
\la{eq:l.neq.0-main}
\ee
The relation of the Matsubara zero-mode $\ell=0$ with 
the zero-frequency retarded correlator requires special care. 
We shall see that their difference can
be expressed in terms of the spectral function.

The spectral function is defined as 
\be
\rho(\omega) = \frac{1}{\pi} \im \tilde G_R(\omega)\,.
\la{eq:rhodef}
\ee
For real $\omega$ and in finite volume, 
$\rho$ is a distribution, namely a discrete sum of delta functions.
However, in the infinite-volume limit of an interacting theory, 
it is typically a smooth function, except possibly at a finite
number of points. The most useful relations among the correlators 
are those that survive the infinite-volume limit.
In the appendix, we show  that 
\be
G_E^{(0)} - \tilde G_R(i\omega_I) =
\lim_{\Lambda\to\infty}
\int_{-\Lambda}^\Lambda \frac{d\omega}{\omega} \, \rho(\omega)\,
 \frac{\omega_I^2}{\omega^2+\omega_I^2}\,,
~~~\forall \omega_I>0.
\la{eq:GR+GE-main}
\ee
In the limit $\omega_I\to 0$, 
only the integration region around the origin contributes 
on the RHS. If the spectral function is finite at the origin,
the RHS is O($\omega_I$), and therefore $\lim_{\epsilon\to0}
\tilde G_R(i\epsilon)=G_E^{(0)}$. On the other hand, if the spectral
function contains a delta-function at the origin, 
$\rho(\omega)/\omega = A\delta(\omega)$, 
$\lim_{\epsilon\to0} \tilde G_R(i\epsilon)=G_E^{(0)}-A$.
We have written \eq(\ref{eq:GR+GE-main}) with an explicit UV cutoff 
to make it clear that one should take the limit $\omega_I\to0$ first.

The connection between the retarded and Euclidean correlators 
is thus precisely established in frequency space. 
In appendix we also rederive the relation between them in coordinate space
(well-known in particular to lattice practitioners), 
\be
G_E(t) = \int_0^\infty d\omega \rho(\omega)
\frac{\cosh\omega(\frac{\beta}{2}-t)}{\sinh\beta\omega/2}\,,
\la{eq:ClatRho}
\ee
which we shall use in section (\ref{sec:num}).

\subsection{The retarded correlator along the imaginary axis}
If we assume for a moment that 
\be
\int_C^\infty \rho(\omega) \frac{d\omega}{\omega},
\qquad (C>0),
\ee
converges in the UV, 
a simple calculation based on the explicit expression
for the spectral density (\ref{eq:rho}), 
or alternatively a contour integration,
then leads to the Kramers-Kronig relation
\be
\int_{-\infty}^{+\infty}d\omega\, \rho(\omega)
\frac{\omega}{\omega^2+\omega_{\rm I}^2}
= \tilde G_R(i\omega_{\rm I})\,,
\qquad\forall \omega_{\rm I}>0.
\la{eq:KK-main}
\ee
In general, it is necessary to perform a subtraction
from $\rho$ to make the dispersion
integral converge in quantum field theory. 
Starting from (\ref{eq:KK-main}), it is easy to 
derive by recursion the following equalities,
\ba
(-1)^n\frac{d^{2n}}{d\omega_{\rm I}^{2n}}
\tilde G_R(i\omega_{\rm I}) &=& 
\int_{-\infty}^{\infty} d\omega \rho^{(2n)}(\omega)\,
\frac{\omega}{\omega^2+\omega_{\rm I}^2}
\\
(-1)^{n+1}\frac{d^{2n+1}}{d\omega_{\rm I}^{2n+1}}
\tilde G_R(i\omega_{\rm I})
&=& 
\int_{-\infty}^\infty d\omega \rho^{(2n+1)}(\omega) 
\frac{\omega_{\rm I}}{\omega^2+\omega_{\rm I}^2}
\stackrel{\omega_I\to0}{=} \pi \rho^{(2n+1)}(\omega=0)\,.
\la{eq:dGR}
\ea
Thus at the origin, the even derivatives of the retarded
correlator along the imaginary axis are given by
integrals over the spectral density, while the odd
derivatives are equal to the corresponding 
derivatives of the spectral function.

\subsection{Remarks on the Euclidean correlator} 
Combining \eq(\ref{eq:KK-main}) with 
\eq(\ref{eq:GR+GE-main}), one obtains
\be
G_E^{(0)}=\int_{-\infty}^\infty d\omega\,
 \frac{\rho(\omega)}{\omega}\,.
\la{eq:GErho}
\ee
Equations (\ref{eq:GR+GE-main}) and (\ref{eq:GErho}) 
show that while 
$G_R(i\omega_{\rm I})$ is insensitive to the presence of 
a $\omega\delta(\omega)$ term in the spectral function 
for any $\omega_{\rm I}>0$, 
$G_E^{(0)}$ definitely receives a contribution from it.

Secondly we remark that the positivity of $\rho(\omega)/\omega$
implies the positivity of certain linear combinations of 
the Euclidean correlator. For instance,
\ba
G_E^{(0)} -G_E^{(\ell)}
&=&\omega_\ell^2
\int_{-\infty}^\infty d\omega\,\frac{\rho(\omega)}{\omega}\,
\frac{1}{\omega^2+\omega_\ell^2}\,,
\la{eq:lincomb1}
\\
3G_E^{(0)} - 4G_E^{(\ell)} + G_E^{(2\ell)}
&=& 12\omega_\ell^4\int_{-\infty}^\infty d\omega
\frac{\rho(\omega)}{\omega}\,
\frac{1}{(\omega^2+\omega_\ell^2)(\omega^2+4\omega_\ell^2)}\,.
\la{eq:lincomb2}
\ea

\subsection{An example}
Here we anticipate the results of section (\ref{sec:bulk})
to illustrate the general formulas of this section.
In Figure (\ref{fig:sketch}), we sketch a possible
form of the retarded correlator for the operator 
$T_{\mu\mu}$ projected onto zero spatial momentum.
In order to deal with a UV finite function,
we subtract the corresponding zero-temperature correlator,
\be
\Delta\tilde G_R(i\omega)= \tilde G_R(i\omega,T) - \tilde G_R(i\omega,T=0)
\ee
and leave the temperature-dependence implicit.
This subtraction is sufficient to make the 
dispersion integral (\ref{eq:KK-main}) converge.
As we shall see in section (\ref{sec:hydro}), the 
spectral function of the operator $\int d^3{\bf x}\, T_{\mu\mu}(x)$
admits a delta function located at the origin, with a weight
given by a thermodynamic quantity. Applying \eq(\ref{eq:GR+GE-main}),
we obtain $\Delta \tilde G_R(i\epsilon)= \Delta G_E^{(0)}-Tc_v(3c_s^2-1)^2$.
Furthermore an exact sum rule fixes $\Delta G^{(0)}_E$
in terms of thermodynamic functions, and  
$\Delta \tilde G_R(i\epsilon)$ turns out to be negative.
Using the operator-product expansion 
(section \ref{sec:OPE}), one can show that 
$\Delta G_E^{(\ell)}$ is negative at large Matsubara frequencies, 
implying the same for $\Delta\tilde G_R(i\omega)$  at large $\omega$
in view of \eq(\ref{eq:l.neq.0-main}).
Finally, \eq(\ref{eq:dGR}) and (\ref{eq:kubo}) show that 
$(d/d\omega)\Delta \tilde G_R(i\omega)$ at the origin 
is proportional to the bulk viscosity,
which has to be positive by the second law of thermodynamics. 
All this information is summarized in \fig(\ref{fig:sketch}).
The behavior at intermediate 
frequencies $\omega$ is not known to us, in principle
$\Delta\tilde G_R(i\omega)$ could even 
change sign in view of the subtraction made. 
It is also not known whether the plotted function 
ever exceeds unity for
$\omega$ a multiple of $2\pi T$.

 \FIGURE[t]{
 \centerline{\includegraphics[width=8.0 cm,angle=-90]{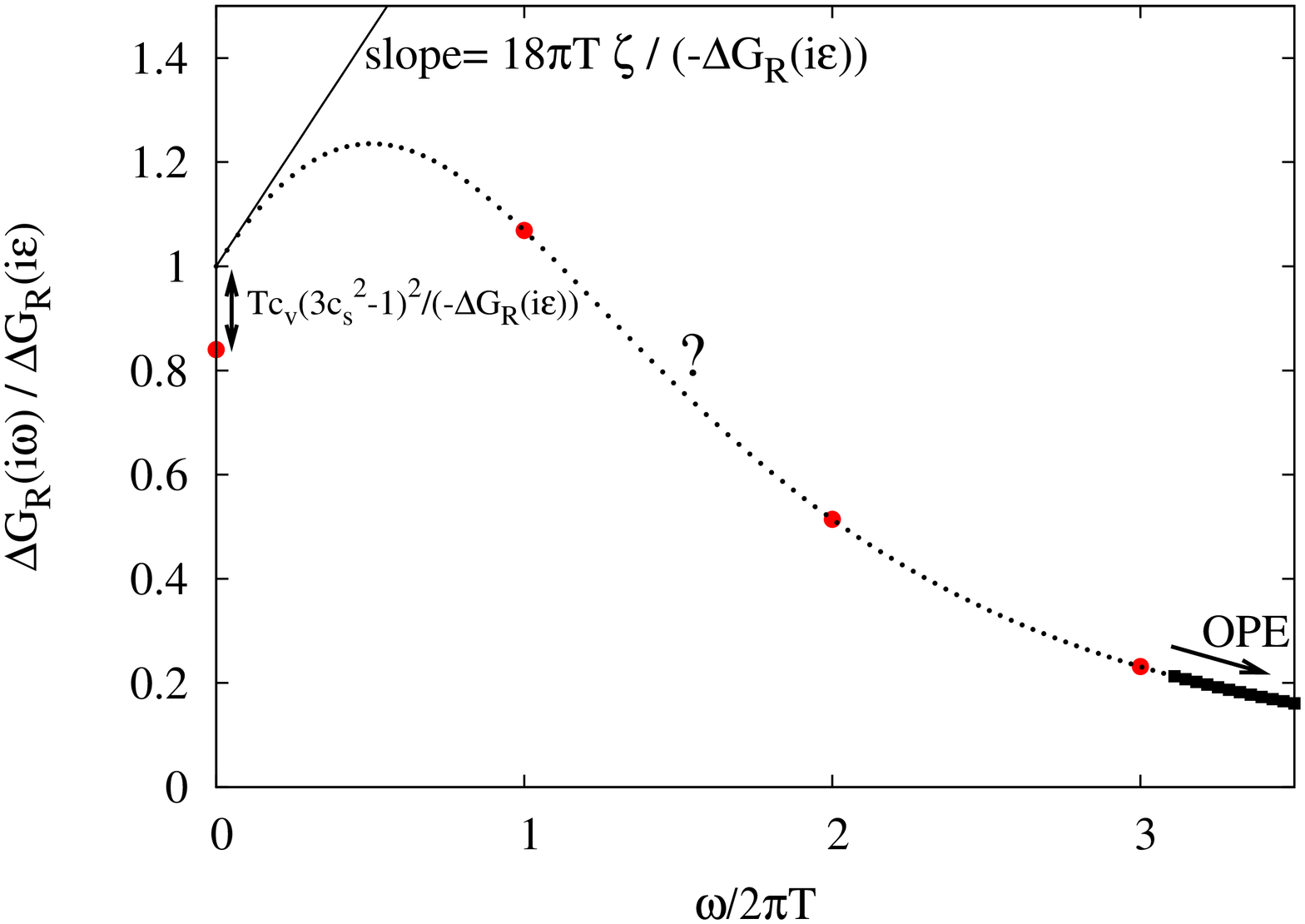}}
 \caption{A sketch of the retarded correlator for the operator 
$L^{-3/2}\int d^3{\bf x}T_{\mu\mu}(x)$ at purely imaginary
frequencies (dotted line). The large dots indicate 
the values of $\Delta G_E^{(\ell)}$, calculable in the Euclidean theory.
Figure (\ref{fig:RHS}) shows that 
 $\Delta \tilde G_R(i\epsilon) = 3(e+p)(1-3c_s^2) -4(e-3p)$
is negative in the deconfined phase.
At high frequencies, $\Delta \tilde G_R$ is constrained by the OPE.
 \label{fig:sketch}
 }
}

In a numerical Euclidean approach, the available information is
$\tilde G_R(i\omega_\ell)$, $\ell=0,1,\dots$
The low-frequency real-time properties 
of the system, such as the bulk viscosity and the associated 
relaxation time,  are determined by the odd derivatives 
of $\tilde G_R(i\omega)$ at the origin, see \eq(\ref{eq:dGR}).
The latter are difficult to determine, because $\omega=0$ is a
non-analytic point of $\tilde G_R$.
In this respect, \eq(\ref{eq:dGR}) shows that 
if the spectral density has a gap $m$,
as it does at zero temperature in a confining theory,
then $\tilde G_R(i\omega)$ 
can be expanded in integer powers of $\omega^2$.
Therefore subtracting the zero-temperature correlator
only affects the even derivatives of 
$\Delta\tilde G_R(i\omega)$ at the origin.

Working in frequency space
has the advantage that the high-frequencies 
are decoupled from the discussion at this stage.
There are, however, technical difficulties 
in determining Euclidean correlators 
in frequency space. Unlike the $t$-dependent correlators,
they are off-shell correlators, and therefore 
require additive counterterms in order to approach the continuum
limit with 
corrections that are suppressed with a power of the lattice
spacing. An example of this complication arises 
in the derivation of sum rules~\cite{Meyer:2007fc}. 
In this special case
the sum rule precisely dictates what the counterterms are.
In general, renormalization conditions have to be imposed
in order to determine them.

Finally we make the observation that a polynomial of degree $n$ 
in $\omega$ going through the $n+1$ points 
$\{\tilde G_R(i\epsilon)\} \cup \{G_E^{(\ell)}\}_{\ell=1}^n$,
has a slope at the origin given by 
\be
-\frac{n!}{\omega_M}
\sum_{\ell=1}^n \frac{(-1)^{\ell}}{\ell!(n-\ell)!} 
\frac{G_E^{(\ell)}-\tilde G_R(i\epsilon)}{\ell}\,.
\ee
When expressed 
in terms of dispersion integrals as in 
\eq(\ref{eq:lincomb1}-\ref{eq:lincomb2}), 
the degree of convergence
in the UV of this expression 
improves with the order of the polynomial: with some 
combinatorics one can show that it is 
asymptotically $\rho(\omega)/\omega^{n+3}$ for $n$ even and 
$\rho(\omega)/\omega^{n+2}$ for $n$ odd.
In particular, the slope at the origin of the $n=2$ polynomial 
is given by
\be
-\frac{1}{2\omega_M} (3\tilde G_R(i\epsilon) - 4G_E^{(1)} + G_E^{(2)})
=
-6\omega_M^3\int_{-\infty}^\infty
 d\omega \,\frac{\rho(\omega)\omega}{\omega^2+\epsilon^2}
\frac{1}{(\omega^2+\omega_M^2)\,(\omega^2+4\omega_M^2)}\,,
\la{eq:slope2}
\ee
where the linear combination (\ref{eq:lincomb2}) appears.
In the specific case of the bulk channel,
the spectral integral (\ref{eq:slope2})
is UV-convergent without any subtraction to $\rho(\omega)$.
Thus the slope of the quadratic polynomial is negative-definite,
just as the actual slope of the retarded correlator 
$\frac{d}{d\omega}\tilde G_R(i\omega)$. One might be tempted
to use it as a Euclidean estimator for the bulk viscosity~$\zeta$.
However, if the spectral function has a transport peak 
at the origin of width 
$\omega_0\lesssim \omega_M$, expression (\ref{eq:slope2}) amounts
parametrically to $\zeta \omega_0/\omega_M$, while
$\frac{d}{d\omega}\tilde G_R(i\omega)\sim \zeta$.
It is clear why that is: a polynomial interpolating between
the Euclidean points is only a good 
approximation to the retarded correlator if the latter 
does not contain a scale $\omega_0\ll \omega_M$
over which it varies significantly.

\subsection{The reconstructed Euclidean correlator}

When analyzing Euclidean correlators obtained
from Monte-Carlo simulations, it is instructive
to study in what way the Euclidean correlators
differ from what they would be if the spectral 
density was unaffected by thermal effects~\cite{Datta:2003ww}:
\be 
G_E^{\rm rec}(t,T;T') {\equiv} \int_0^\infty d\omega \rho(\omega,T')
\frac{\cosh\omega(\frac{\beta}{2}-t)}{\sinh \omega \beta/2} \,.
\la{eq:Grec1-main}
\ee
The identity ($0\leq t\leq \beta$)
\be
\frac{\cosh\omega(\frac{\beta}{2}-t)}{\sinh \omega \beta/2}
= \sum_{m\in{\bf Z}} e^{-\omega|t+m\beta|}
\ee
allows us to derive in particular the exact relations
\ba
G_E^{\rm rec}(t,T;0) &=& \sum_{m\in{\bf Z}} G_E(|t+m\beta|,T=0)\,,
\la{eq:Grec2}
\\
G_E^{\rm rec}(t,T;{\txts\half} T) &=& 
G_E(t,{\txts\half} T) + G_E(\beta-t, {\txts\half} T)\,.
\ea
These expressions are useful in practice when 
one is interested in a difference of spectral functions.
We will exploit \eq(\ref{eq:Grec2}).
Even if the zero-temperature 
Euclidean correlator can usually not 
be determined for arbitrarily large separations,
in that regime it is completely dominated by the lightest state 
in the channel, so that its functional form is 
known and completely determined by one matrix element
and an energy. 
The contribution to $G_E^{\rm rec}$
of a single state of energy $E$ is easily obtained,
\be
\delta G_E^{\rm rec}(t,T) \sim
\frac{\cosh E(\beta/2-t)}{\sinh \beta E/2}\,.
\ee
Secondly, the correlator falls off
very rapidly, so that the terms with large values 
of $m$ in \eq(\ref{eq:Grec2}) make a very small contribution.

\section{The bulk channel in QCD\la{sec:bulk}}
In this section the goal is to provide the theoretical background
for the analysis of the bulk channel in SU($N$) gauge theory.
Thus we consider the operator
\be
{\cal O}_\theta(t)=\frac{1}{L^{3/2}}\int d^3{\bf x}~\theta(t,{\bf x})\,,
\qquad \theta(x)\equiv 
T_{\mu\mu}(x)={\txts\frac{\beta(g)}{2g}}G_{\mu\nu}^a(x)G_{\mu\nu}^a(x)\,.
\ee
Here $\beta(g)$ is the beta function, 
$\beta(g)=-bg^3+\dots$, $b=\frac{11N}{3(4\pi)^2}$,
and $T_{\mu\nu}(x)$ is the energy-momentum tensor.
Its spectral function $\rho_\theta(\omega,T)$ and the difference 
\be
\Delta\rho_\theta(\omega,T) \equiv 
\rho_\theta(\omega,T)   -\rho_\theta(\omega,0)
\ee
play an important role in the following.
The spectral function determines the bulk viscosity through
the Kubo formula
\be
\zeta(T) = \frac{\pi}{9} \lim_{\omega\to0}
        \frac{\rho_\theta(\omega)}{\omega}.
\la{eq:kubo}
\ee

In what follows 
we review the behavior that hydrodynamics predicts
for the low-frequency part of the
associated spectral function $\rho_\theta$.
It turns out that it contains a term $\omega\delta(\omega)$.
We then define an `auxiliary' operator whose spectral function
coincides with $\rho_\theta$ except for not having the 
delta function at the origin. 
Next we present the asymptotic large-$\omega$ behavior 
of $\rho_\theta$ based on the operator-product expansion.
Finally, the spectral function of the auxiliary operator 
obeys a sum rule given in section (\ref{sec:OPE}).

\subsection{Low-frequency prediction from hydrodynamics\la{sec:hydro}}
In this section we consider more general connected 
correlation functions of the energy-momentum tensor,
\be
G_{E,\mu\nu\rho\sigma}(t,{\bf k}) = 
\int d^3{\bf x} \, e^{i{\bf k\cdot x}}\,
\< T_{\mu\nu}(t,{\bf x})\,T_{\rho\sigma}(0)\>\,,
\ee
as well as their associated spectral function
 $\rho_{\mu\nu\rho\sigma}(\omega,{\bf k},T)$ via (\ref{eq:ClatRho}).
The spectral function for $T_{33}$ (assuming the 
spatial momentum ${\bf k}$ is pointing in the direction $\hat 3$)
reads~\cite{Teaney:2006nc}
\be
\frac{\rho_{33,33}(\omega,{\bf k},T)}{\omega} = \frac{e+p}{\pi} 
\frac{\Gamma_s \omega^4}{(\omega^2-c_s^2k^2)^2+(\Gamma_s \omega k^2)^2}.
\la{eq:sound}
\ee
Now one shoud recall that 
for any smooth function such that $\int_{-\infty}^\infty f(x)dx=1$,
$\frac{1}{\epsilon}f(x/\epsilon)$ provides a representation
of the delta function. Here we will exploit this fact for 
the function
\be
f(x) = \frac{1}{\pi(b^2-1)}\left[
\frac{x^4}{(x^2-b^2)^2+x^2}-1\right]
\ee
and $b=c_s/(\Gamma_s k)$ and $\epsilon = \Gamma_s k^2$.
In this way one finds that, in the sense of distributions, 
\be
\lim_{{\bf k}\to0}
\frac{\rho_{33,33}(\omega,{\bf k},T)}{\omega} =
\frac{e+p}{\pi} \Gamma_s  + (e+p)c_s^2\delta(\omega) + \dots
\ee
This is in contrast with what is obtained if one 
simply sets ${\bf k}=0$ in \eq(\ref{eq:sound}), in which 
case one misses the delta function.
In the shear channel on the other hand, 
\be
\frac{\rho_{13,13}(\omega,{\bf k},T)}{\omega}
= \frac{\eta}{\pi} \frac{\omega^2}
       {\omega^2+\left(\frac{\eta k^2}{e+p}\right)^2}
\stackrel{{\bf k}\to0}{\sim} \frac{\eta}{\pi} 
-\frac{\eta^2 k^2}{e+p}\delta(\omega)
\ee
and the delta function is suppressed by a power of ${\bf k}^2$.
Thus combining the shear and sound channels,
\be
\frac{1}{9}\frac{\rho_{ii,kk}(\omega,{\bf 0},T)}{\omega}
= \frac{\rho_{33,33}(\omega,{\bf 0},T)}{\omega}
 - \frac{4}{3}  \frac{\rho_{13,13}(\omega,{\bf 0},T)}{\omega}
= \frac{\zeta}{\pi}+(e+p)c_s^2\delta(\omega)+\dots
\ee
Using the exact relations
\ba
\rho_{00,00}(\omega,{\bf 0},T) 
&=& \frac{e+p}{c_s^2} \omega \delta(\omega)\,,
\\
\rho_{00,kk}(\omega,{\bf 0},T) 
&=& -3(e+p) \omega \delta(\omega)\,,
\ea
one finds
\be
\frac{\rho_\theta(\omega,{\bf 0},T)}{\omega}
\equiv
\frac{\rho_{\mu\mu,\nu\nu}(\omega,{\bf 0},T)}{\omega}
= \frac{9\zeta}{\pi} 
+ \frac{e+p}{c_s^2}(3c_s^2-1)^2\delta(\omega)
+\dots
\la{eq:separ}
\ee
When one constrains the spectral function using 
sum rules and Euclidean correlators, one is always
dealing with integrals of the spectral function
over all frequencies. Therefore it is useful
to subtract the contribution of the delta function
in (\ref{eq:separ}).
One way to achieve this is to work with the operator
\be
{\cal O}_\star = L^{-3/2}
\int d^3{\bf x}\,\left(T_{\mu\mu}(x)+(3c_s^2-1)T_{00}(x)\right)\,.
\la{eq:Ostar}
\ee
Its spectral function, which we denote by $\rho_\star$,
satisfies
\be
\rho_{\theta}(\omega,{\bf 0},T)
= \rho_{\star}(\omega,{\bf 0},T)
+ \frac{e+p}{c_s^2}(3c_s^2-1)^2\,\omega\,\delta(\omega)
\la{eq:separ2}
\ee
and, in view of \eq(\ref{eq:separ}), 
is free of the delta function at the origin.
At equilibrium, recalling that $c_s^2=\frac{\partial p}{\partial e}$,
\be
-\frac{1}{3}\<{\cal O}_{\star}\> =  
p-{\txts\frac{\partial p}{\partial e}} e
\ee
and therefore its linear fluctuations are not correlated
with those of the total energy\footnote{Recall
that we are working in the canonical ensemble. I thank G.D.~Moore
for pointing this out to me.}.
This is the physical reason why a delta function at the origin 
is absent in its spectral function.

\subsection{Large-frequency behavior 
from the operator product expansion\la{sec:OPE}}
In section (\ref{sec:hydro}) 
we have analyzed the low-frequency behavior of 
the bulk spectral function using the general
hydrodynamic predictions. We now turn to an analysis 
of its asymptotic high-frequency behavior. The leading 
$\rho_\theta\sim \alpha_s^2(\omega)\omega^4$ 
behavior was given for instance in~\cite{Meyer:2008gt}.
In the OPE language, this contribution corresponds to 
the insertion of the unit operator, and it is therefore
temperature independent. Because it makes a large contribution
to the coordinate-space Euclidean correlator, it is 
best to remove it by subtracting the vacuum correlator
from the thermal correlator. We therefore need to know what
the leading behavior of the subtracted correlator 
and of the corresponding subtracted spectral function is.

The operator product expansion of local gluonic operators
was calculated a long time ago in the context of 
QCD sum rules~\cite{Novikov:1979va}, and 
the results of more recent calculations are
summarized in~\cite{Forkel:2003mk}.
In many cases, the Wilson coefficients are known at 
next-to-leading order.
However, only the Lorentz scalar operators were kept on
the right-hand side of the OPE, since the correlator
was evaluated in the vacuum. Since we want to apply 
the OPE to a system at finite-temperature, the traceless
part of the energy-momentum tensor, for instance,
can also contribute.
The OPE results therefore need to be generalized to 
the finite-temperature context~\cite{CaronHuot:2009ns}.
We have reproduced the result of~\cite{CaronHuot:2009ns}
for the bulk channel using the background field method 
combined with the Fock-Schwinger gauge~\cite{Novikov:1983gd},
\be
\la{eq:OPE}
\Delta\rho_{\theta}(\omega,{\bf 0},T) 
\stackrel{\omega\to\infty}{\sim}
-\frac{1}{2}\left(\frac{11N\alpha_s(\omega)}{6\pi}\right)^2 (e-3p)
-\frac{3}{4} \left(\frac{11N\alpha_s(\omega)}{6\pi}\right)^3 (e+p)\,.
\ee
We remark that while the two-point function $\Delta\tilde G_R(i\omega)$ 
of the operator $T_{\mu\nu}-\frac{1}{4}\delta_{\mu\nu}\theta$
contains a logarithmic divergence
(which manifests itself in the Wilson coefficient of $\theta$
being O($1/g_0^2\sim \log 1/a$) in lattice regularization~\cite{Meyer:2007fc}),
$\Delta\tilde G_R(i\omega)$ is UV-finite for the operator $\theta$.
Therefore, assuming the validity of the operator-product expansion,
the only relevant scale in its Wilson coefficients is $\omega$, 
and the dispersion relation 
(\ref{eq:KK-main}) then leads to \eq(\ref{eq:OPE})~\cite{Huang:1994vd}.
In section (\ref{sec:num}),
we will use this information to constrain the form 
of the spectral function at large frequencies.

\subsection{The Bulk Sum Rule}
The idea of constraining the form of finite-temperature 
spectral functions with sum rules is not 
new~\cite{Kapusta:1993hq,Huang:1994vd,Kharzeev:2007wb}.
Here we focus on the bulk channel.
In Euclidean space, the following sum rule 
holds~\cite{Ellis:1998kj},
\be
G_{E,\theta}^{(0)}(T) -G_{E,\theta}^{(0)}(T=0)
= T^5\frac{\partial}{\partial T} \frac{e-3p}{T^4}\,,
\la{eq:srE}
\ee
where $e$ and $p$ are the energy density and pressure of the 
finite-temperature system. 

We now convert identity (\ref{eq:srE}) into a
sum rule for the bulk spectral function
with the goal to constrain the latter.
Combining equations 
(\ref{eq:srE})  and (\ref{eq:GErho}), we obtain
\be
\int_{-\infty}^\infty  \frac{d\omega}{\omega} 
[\rho_{\theta}(\omega,T) -\rho_\theta(\omega,0)] = 
T^5 \partial_T \frac{e-3p}{T^4} \,.
\la{eq:srtheta}
\ee
Next, we separate the contribution
of the $\omega\delta(\omega)$ term in $\rho_\theta$
from the rest of the spectral function.
In other words, we want to turn \eq(\ref{eq:srtheta})
into a sum rule for $\rho_\star$, the spectral function
of the operator ${\cal O}_\star$ (\ref{eq:Ostar}),
which is smooth at the origin.
Using \eq(\ref{eq:separ2}),  one finds
\be
2\int_0^\infty \frac{d\omega}{\omega}
[\rho_{\star}(\omega,T)
 -\rho_{\star}(\omega,0)]
= 3(1-3c_s^2)(e+p) -4(e-3p)\,.
\la{eq:bsr1}
\ee
A few remarks are in order.
It would have been less straightforward
to directly derive a sum rule for $\rho_\star$,
because the Euclidean correlator of ${\cal O}_\star$
contains a short-distance singularity.
We remark that \eq(\ref{eq:srtheta}) was 
already obtained in~\cite{Kharzeev:2007wb}, but the 
presence of the $\omega\delta(\omega)$ term 
in $\rho_\theta$ was missed.
In~\cite{Romatschke:2009ng}, a finite spatial momentum ${\bf k}$
was used as an infrared regulator and the contribution of the 
zero-frequency delta function correctly identified 
for the first time.
If one insists on expressing \eq(\ref{eq:bsr1}) 
in terms of $\rho_\theta$, one should strictly
write $\lim_{\epsilon\to0} \int_{-\infty}^\infty
\frac{\omega \,d\omega}{\omega^2+\epsilon^2}
 \Delta\rho_\theta(\omega,T) = 3(1-3c_s^2)(e+p) -4(e-3p)$.

\section{Numerical results from the lattice\la{sec:num}}
In this section we describe the lattice setup and the 
numerical results obtained by Monte-Carlo simulations.
We employ the isotropic Wilson action~\cite{Wilson:1974sk},
\be
 S_{\rm g} =  \frac{1}{g_0^2} \sum_{x,\mu\neq\nu} \tr\{1-P_{\mu\nu}(x)\}\,,
\la{eq:Sg}
\ee
where the `plaquette' $P_{\mu\nu}$ is the product of four link 
variables $U_\mu(x)\!\in\,$SU(3) around an elementary cell in the $(\mu,\nu)$
plane. As a discretization of $L^{3/2}{\cal O}_\theta$, we use
\be
a^3\sum_{\bf x}  \Theta(x) 
= \frac{-2}{a} \frac{dg_0^{-2}}{d\log a} 
\sum_{\bf x} \re\tr 
 \Big\{ {\txts\sum_{k}}  P_{0k}(x) 
+  {\txts\sum_{k<l}} \half[P_{kl}(x)+P_{kl}(x+a\hat0) ] \Big\},
\la{eq:Theta}
\ee
up to an irrelevant additive constant (its contribution vanishes
in connected correlation functions).
As a local update algorithm, we use the standard combination of heatbath and
over-relaxation~\cite{Creutz:1980zw,Cabibbo:1982zn,Kennedy:1985nu,Fabricius:1984wp}
sweeps in a ratio increasing from
3 to 5  as the lattice spacing is decreased.
The multi-level algorithm is used to reduce 
the variance of the correlator~\cite{Meyer:2002cd,Meyer:2003hy}.
To set the scale and to evaluate the lattice beta function
$dg_0^{-2}/d\log a$ we use the parametrization of the 
Sommer scale $r_0\approx0.5$fm given in~\cite{Necco:2001xg}.
We denote the mass gap by $m$, given by 
the lightest scalar glueball, whose mass is 
$3.96(5)/r_0=m=5.30(8)T_c$.
Finally we relate the notation used in the rest of this section
to the general definitions of sections 
(\ref{sec:review}) and (\ref{sec:bulk}),
\ba
C_\theta(t,T)\equiv \beta^{5} G_{E,\theta}(t),
&\qquad&
C_\star(t,T)\equiv \beta^{5} G_{E,\star}(t),
\\
C^{\rm rec}_\star(t,T;T_0)
\equiv \beta^{5} G^{\rm rec}_{E,\star}(t,T;T_0)\,,
&\qquad&
C^{\rm rec}_\star(t,T)\equiv C^{\rm rec}_\star(t,T;0)\,,
\\
\Delta\rho_\star(\omega,T) = \rho_\star(\omega,T)-\rho_\star(\omega,0)\,.
&&
\ea
The quantities with a $\star$ subscript are associated with the operator
${\cal O}_\star$ defined in \eq(\ref{eq:Ostar}).
The Euclidean correlators $C_{\dots}$ are dimensionless,
renormalization-group invariant quantities.
The goal of this section is to constrain the function
$\Delta\rho_\star$, the main results being 
\eq(\ref{eq:Cmid}) and \eq(\ref{eq:bsr}).

We start by describing a method we have used
to evaluate the right-hand side
of the bulk sum rule (\ref{eq:bsr1}).
We also evaluate the `condensates' appearing in the OPE
(\ref{eq:OPE}) of the bulk correlator.
We obtain $C_\star$ at finite temperature by subtracting
the contribution of the $\delta$ function.
We then describe the properties 
of the correlator at temperatures deep in the confined phase. 
This correlator is subsequently used to produce a 
`reconstructed' correlator based on (\ref{eq:Grec2}). 
From then on we are directly
dealing with integrals of the subtracted spectral function,
$\Delta\rho_\star$. 
We reuse lattice data presented in~\cite{Meyer:2007dy},
and combine it with the bulk sum rule to determine
the gross features of the bulk spectral function.

\subsection{Right-hand side of the bulk sum rule}
 \FIGURE[t]{
 \centerline{\includegraphics[width=8.0 cm,angle=-90]{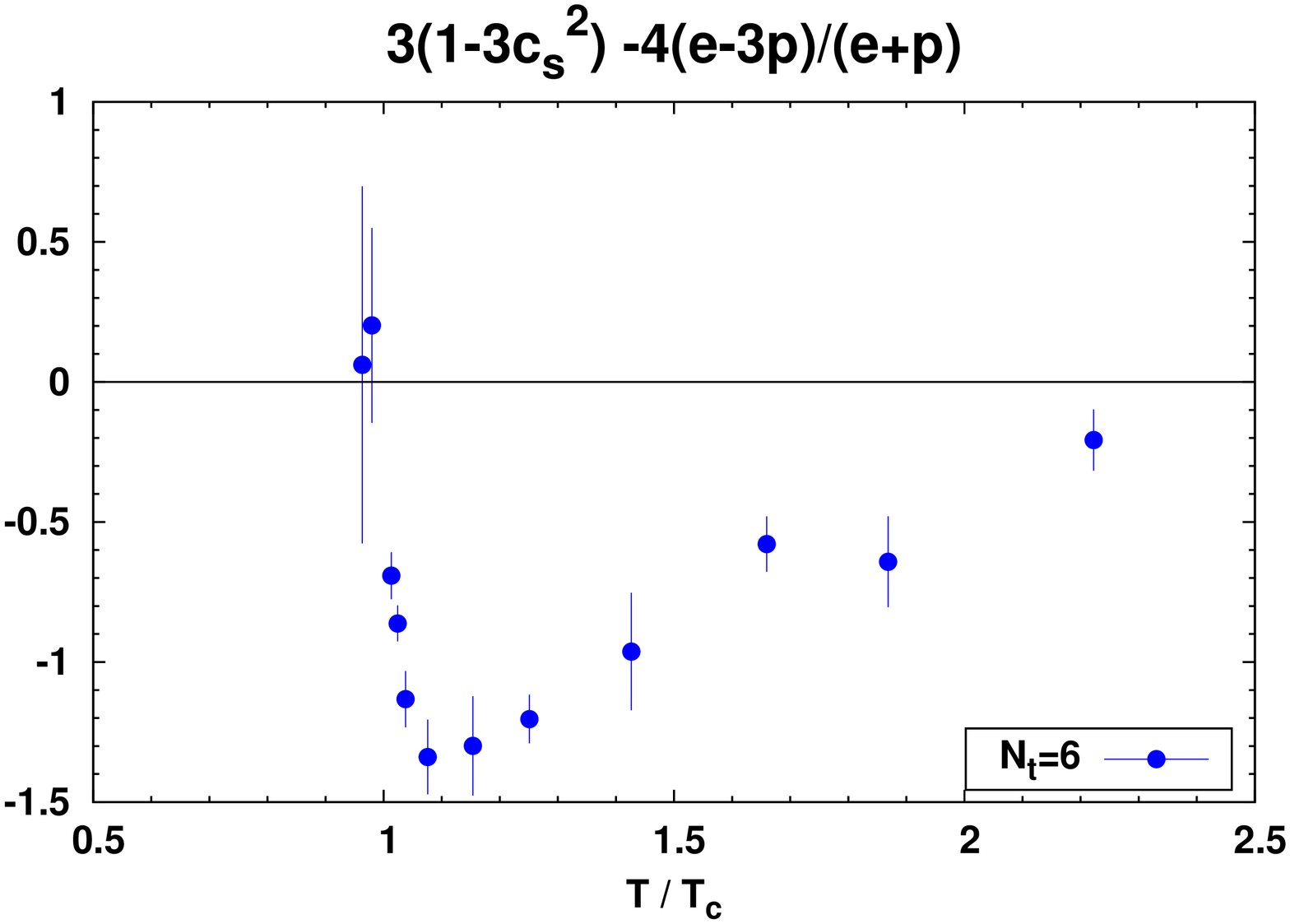}}
\vspace{-0.4cm}

 \caption{The right-hand side of the bulk sum rule as a function of temperature.
  The spatial volume in the finite-temperature simulations is typically $28^3$.}
 \label{fig:RHS}
 }

When using the bulk sum rule (\ref{eq:bsr1}),
the question arises, how to best evaluate 
the combination of thermodynamic quantities 
appearing on the right-hand side of the equation.
Note that the enthalpy $e+p$ and the conformality measure
$e-3p$ are relatively straightforward 
to obtain from the expectation value of the electric and 
magnetic field strengths. What is then needed is a reliable
way to obtain the velocity of sound $c_s^2$, or alternatively,
the specific heat $c_v$.

The Euclidean sum rule (\ref{eq:srE}) 
can be derived on the lattice.
Here we will use that lattice sum rule~\cite{Meyer:2007fc}
to determine the velocity of sound  via
\be
\frac{1}{c_s^2}-3 = \left(4-\frac{d^2g_0^{-2}/d(\log a)^2}{dg_0^{-2}/d\log a}\right) \frac{e-3p}{e+p}
+\frac{a^4}{e+p} \sum_x \<\Theta(x) \Theta(0)\>^{\rm conn}_{T-0}\,,
\la{eq:latsumrule}
\ee
where the discretization $\Theta$ of the trace anomaly operator
is given by \eq(\ref{eq:Theta}).
We used the leading perturbative expression for the expression
inside the large brackets, since based on that estimate 
it only departs from 4 by 0.008 or less.

The right-hand side of the bulk sum rule (\ref{eq:bsr1}), as calculated 
on $\Nt=6$ lattices, is displayed 
in units of $e+p$ in \fig(\ref{fig:RHS}) as a function of temperature.
It is negative for a significant range of temperatures 
above $T_c$, and its leading behavior in perturbation theory 
is~\cite{Kajantie:2002wa}
\be
3(1-3c_s^2) - 4\frac{e-3p}{e+p} = 
-10\cdot \left({\txts\frac{11}{3}}\right)^2\cdot \left(\frac{\alpha_sN}{4\pi}\right)^3
+\dots
\la{eq:PT}
\ee
It is therefore likely to be negative throughout the deconfined phase.
We also note that for $\alpha_s=0.25$, the right-hand side of 
(\ref{eq:PT}) amounts to about -0.03, 
a very small value compared to those appearing in \fig(\ref{fig:RHS}).

\subsection{Evaluation of the correlator $C_\star(t,T)$}
 \FIGURE[t]{
\centerline{\includegraphics[width=6.0 cm,angle=-90]{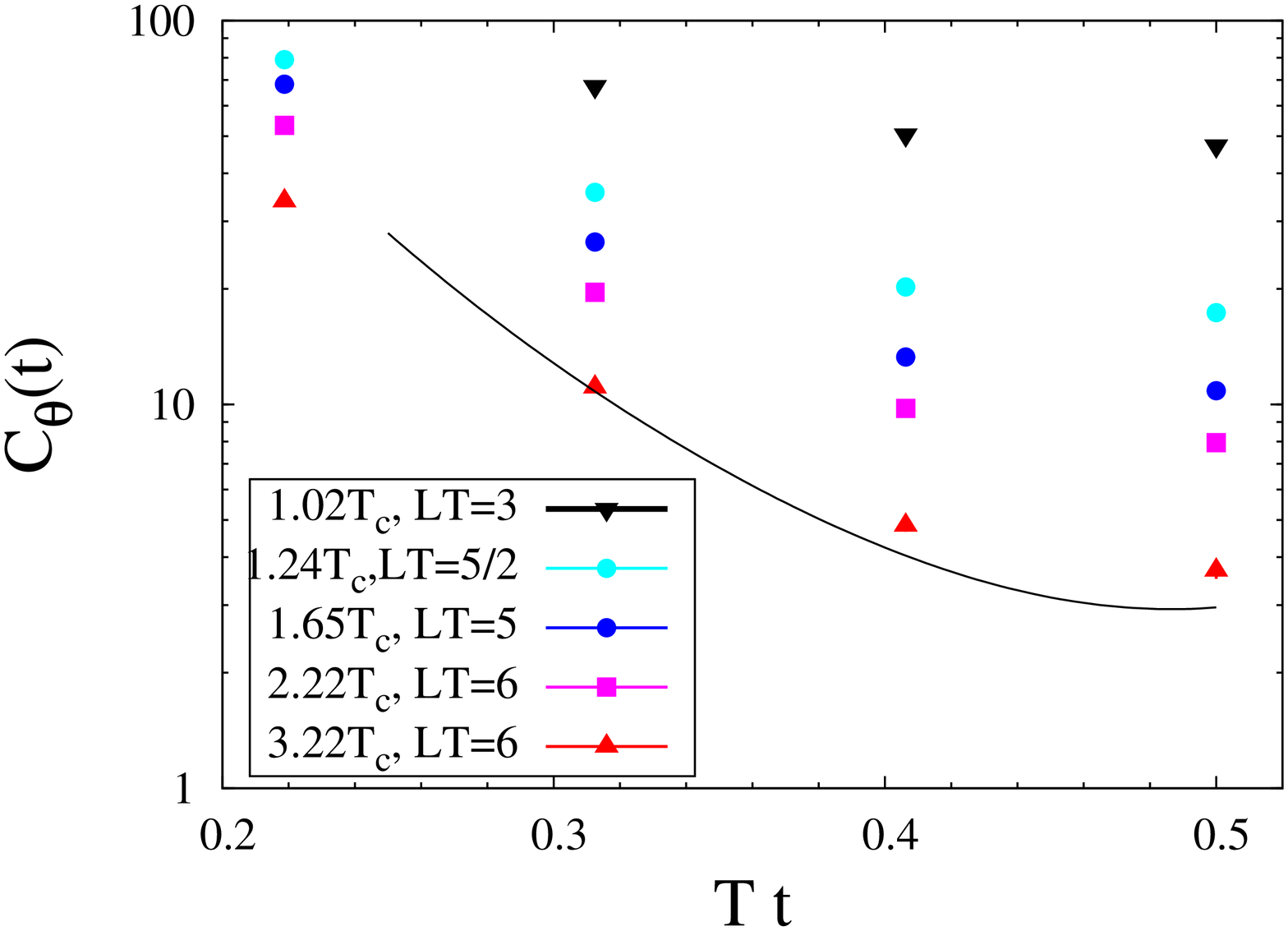}
\includegraphics[width=6.0 cm,angle=-90]{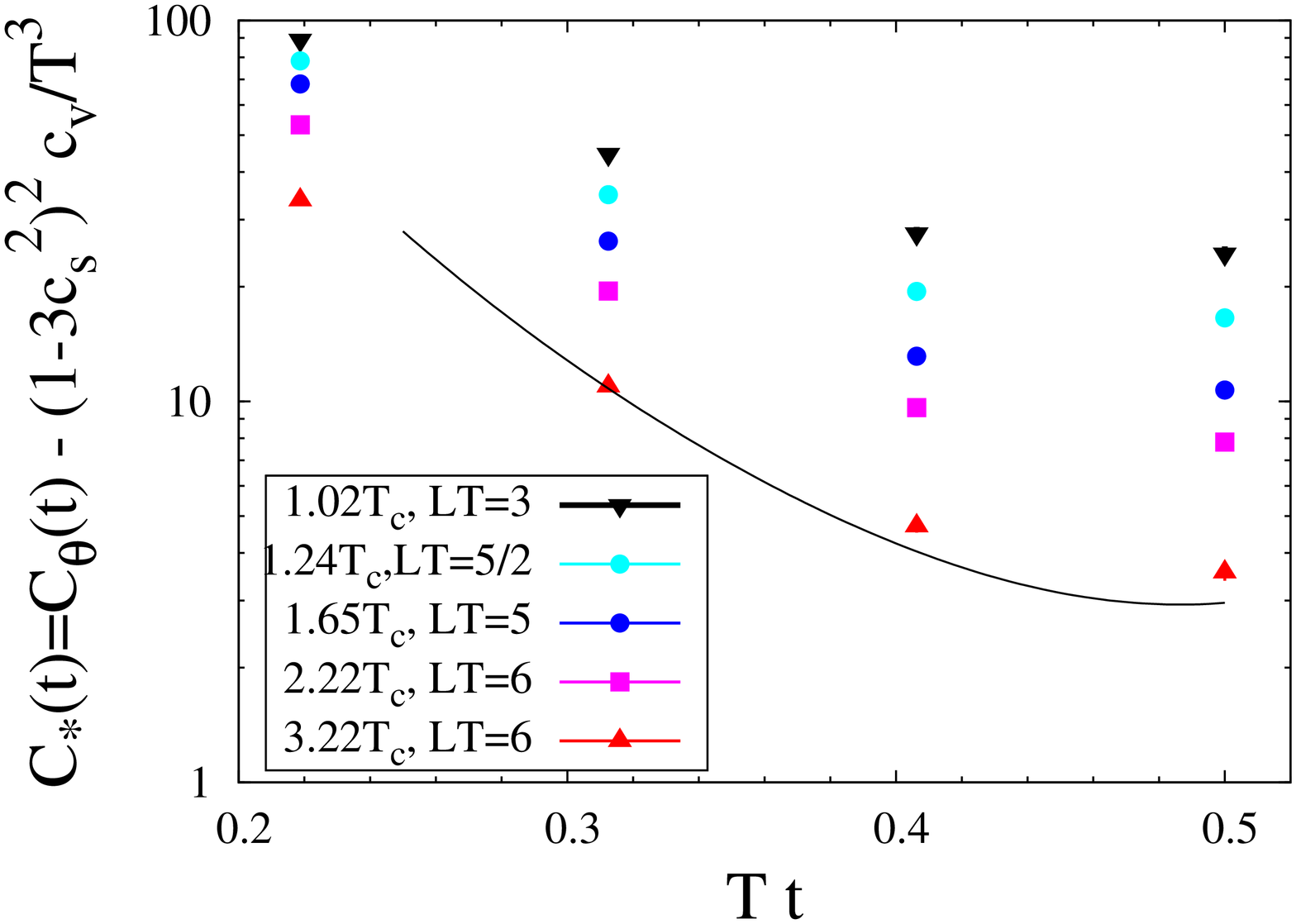}
}
\caption{The treelevel-improved $C_\theta$ 
correlator for several temperatures.
On the right panel, $\frac{c_v}{T^3}\,(1-3c_s^2)^2$ has been subtracted.
}
\label{fig:bulk-corr}
}

Since we want to work with the correlation functions of the 
operator ${\cal O}_\star$, we have to evaluate its Euclidean
correlation function $C_{\star}(t)$.
Lattice perturbation theory~\cite{Meyer:2009vj} and numerical 
studies~\cite{Meyer:2007dy} show that it is advantageous to work 
with the operator ${\cal O}_\theta$.
To convert its correlation function to the 
correlation function $C_\star$, 
we use the equation (see \ref{eq:separ2} and 
\ref{eq:ClatRho})
\be
C_\theta(t,T) = C_\star(t,T) + \frac{e+p}{T^4c_s^2}(3c_s^2-1)^2\,.
\la{eq:Cstar}
\ee
Here too a combination of thermodynamic quantities appears.
We have obtained it by the same method
as the right-hand side of the bulk sum rule, see the previous subsection.
Close to the phase transition where $c_s^2$ is very small,
the result is rather sensitive to the value of $c_s^2$, since it 
appears in the denominator.
That is the main reason we opted for a direct 
calculation of $c_s^2$ from the lattice sum rule (\ref{eq:latsumrule}), 
rather than parametrizing 
the pressure and obtaining $c_s^2$ by taking derivatives. 
In this way we avoid the dependence on a parametrization ansatz, 
which can be significant  in a region  where 
the derivatives are very large.
On the other hand, at high temperatures
\be
\frac{1}{c_s^2}(3c_s^2-1)^2 
= 100\cdot\left({\txts\frac{11}{3}}\right)^2\cdot \left(\frac{N\alpha_s}{4\pi}\right)^4
+\dots
\la{eq:shift}
\ee
and the difference between $C_\theta$ and $C_\star$ is parametrically 
negligible, since both are O($\alpha_s^2$).

The resulting 
correlators are displayed in \fig(\ref{fig:bulk-corr}).
The lattice data for $C_\theta$ is the same as in~\cite{Meyer:2007dy}.
In particular, the lattice spacing is given by $1/aT=8$.
At high temperatures, there is very little numerical difference 
between them, confirming the parametric expectation (\ref{eq:shift}).
On the other hand 
we observe that the difference between $C_\theta$ and $C_\star$ 
becomes appreciable very close to $T_c$. At that point,
$C_\star$ is significantly smaller than $C_\theta$.

\subsection{Low-temperature correlators}

\FIGURE[t]{
\centerline{\includegraphics[width=6.0 cm,angle=-90]{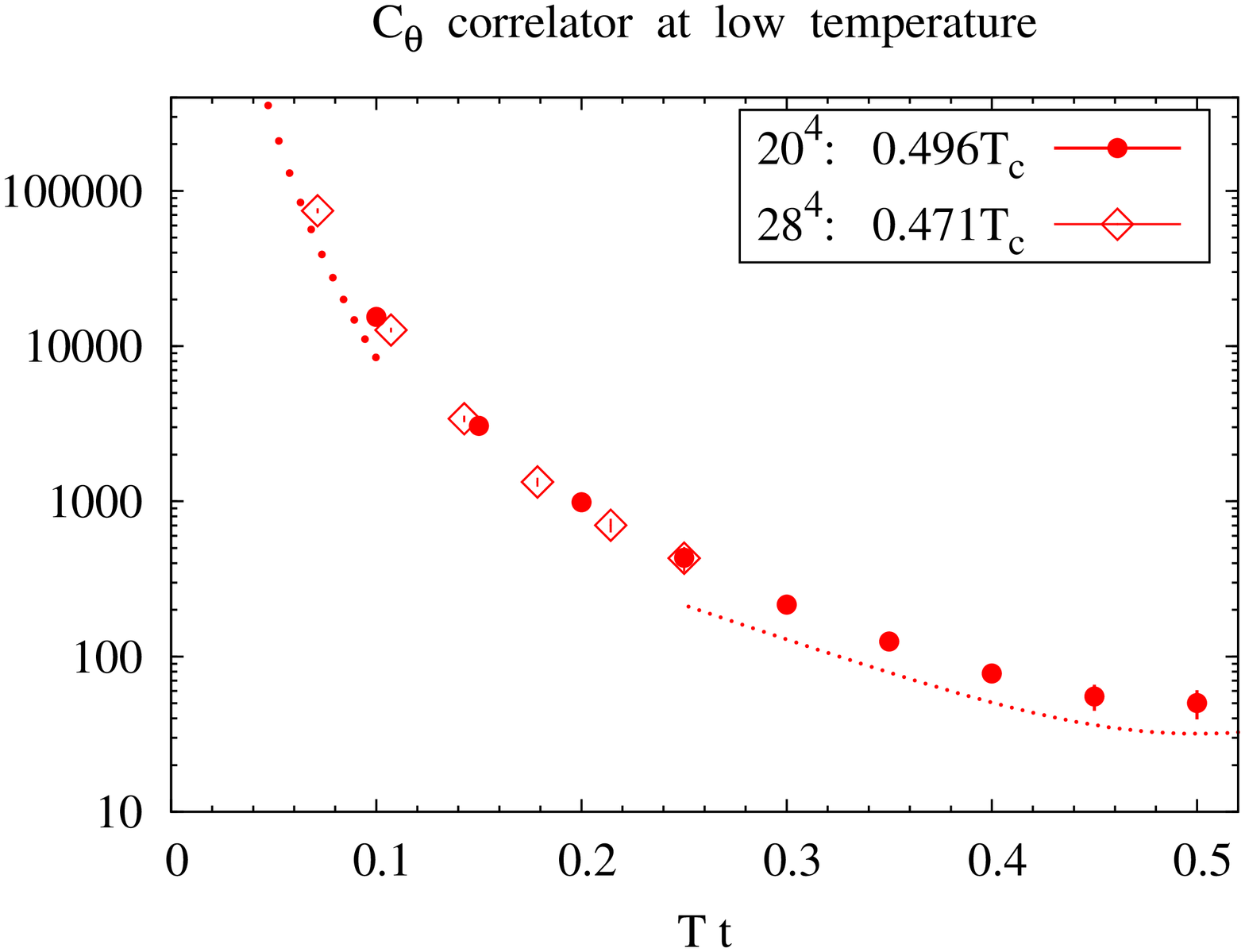}
\includegraphics[width=6.0 cm,angle=-90]{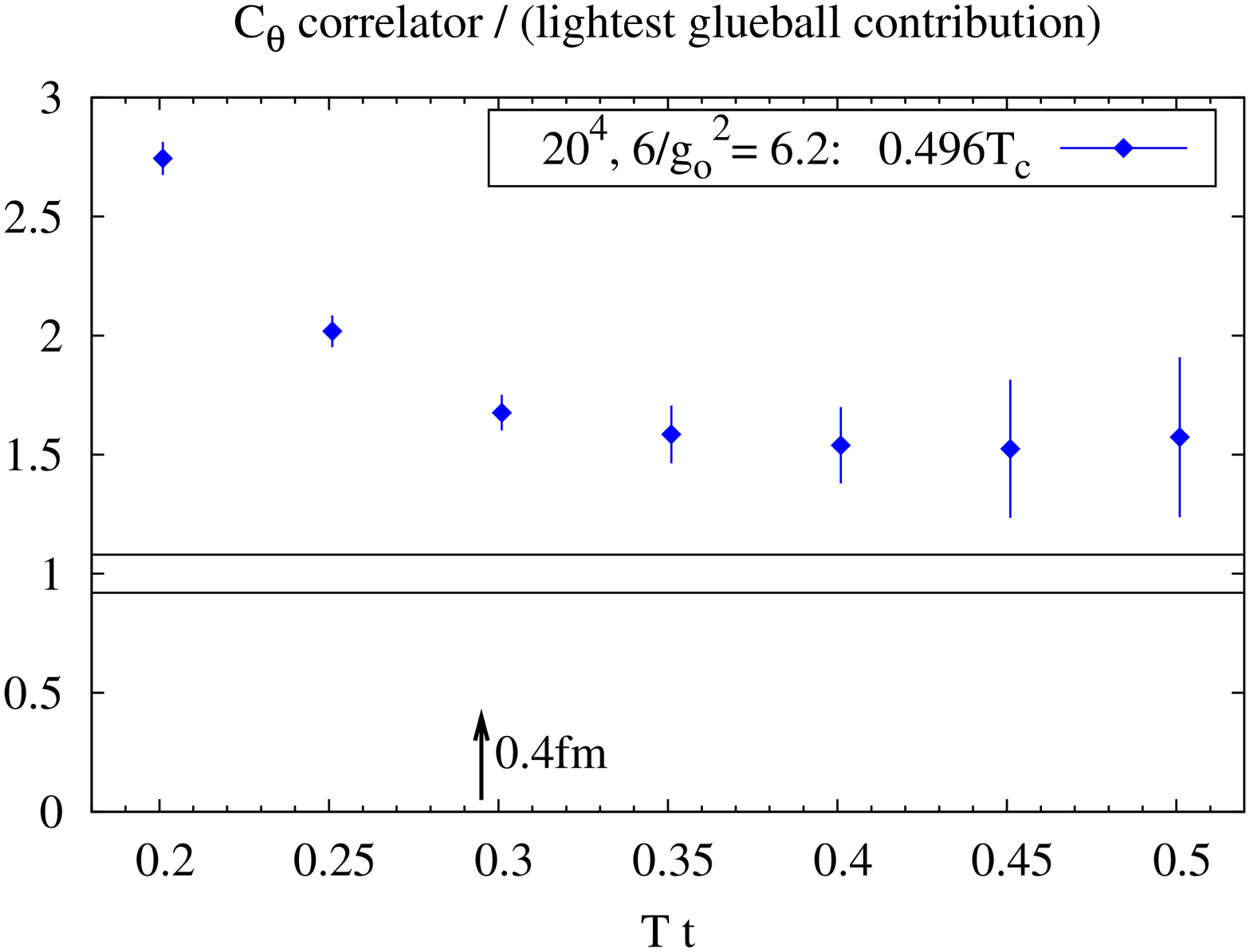}
}
\caption{The $\theta$ correlator at about half the deconfining temperature.
The leading-order perturbative prediction, with $\alpha_s=0.30$, appears 
at small $t$. The contribution of the lightest scalar glueball
is given at larger $t$.
On the right panel, the correlator has been divided by the contribution of 
the lightest scalar glueball. Treelevel improvement has not been applied.
}
\label{fig:corr-zerotemp}
}

The $C_\theta$ correlator at about $T_c/2$ is displayed 
in \fig(\ref{fig:corr-zerotemp}).
The agreement between the data obtained at two different 
lattice spacings suggests that the cutoff effects 
on the correlators are small.
The dotted curve at small separation, which represents the leading
perturbative prediction for the choice $\alpha_s=0.30$,
shows that by comparison 
the non-perturbative  correlator is rather large.
The right panel shows that starting at about 0.4fm,
the lightest glueball represents
a large contribution of the full correlator.
The data is not accurate enough to tell how well 
that state saturates the correlator beyon 0.6fm.

Although the finite size of the lattice implies that 
the temperature is about $T_c/2$, rather than zero,
it is very likely that the spectral function is 
already close to the zero-temperature spectral function,
albeit in finite volume.
This statement can be justified as follows.
The confined phase can be thought of as a dilute gas
of glueballs propagating freely, up to occasional collisions.
Evidence in favor of this picture comes from a precision
thermodynamic calculation in the 
confined phase~\cite{Meyer:2009tq}.
Thus, to leading order at low temperatures, 
\ba
s(\beta) &\approx& m\beta
\left(\frac{m}{2\pi\beta}\right)^{3/2} \, \exp(-\beta m)
\la{eq:lowTentropy}
\\
c_v &\approx &  (\beta m) s(\beta)\,,
\ea
where $m$ is the mass gap.
Based on these approximate expressions,
we find that the contribution of the exact 
delta function to the correlator $C_\theta$ at $t=\beta/2$ is 
\be
\left[C_\theta(t,T)-C_\star(t,T)\right]_{t=\beta/2} \simeq 
(\beta m)^2 \left(\frac{m\beta}{2\pi}\right)^{3/2} e^{-\beta m}\,.
\ee
Numerically, at $T=T_c/2$ this evaluates to about $0.006$, 
while $C_\theta(T,\beta/2)=50(11)$. 
The area under the transport peak can be estimated
using kinetic theory. The general expression is given 
in~\cite{Teaney:2006nc},
\be
\int_{-\Lambda}^\Lambda \frac{\rho_\star(T,\omega)}{T^4}
 \frac{d\omega}{\omega}
= 3\frac{s}{T^3} \<v_{\bf p}^2\> \simeq
15 \left(\frac{m\beta}{2\pi}\right)^{3/2} e^{-\beta m}\,,
\ee
where $\Lambda$ is much larger than the inverse relaxation time,
but lower than the thermal scale.
At low temperatures this contribution is thus parametrically 
smaller than  the contribution of the exact delta function,
and it is indeed less than $15\%$ of the latter at $T_c/2$.

In summary, there are strong reasons to believe that 
the bulk spectral function at $T=T_c/2$ is close to the 
zero-temperature spectral function. Therefore it is legitimate 
to use these Euclidean correlators to obtain the 
reconstructed correlators at temperatures above $T_c$, where 
they can be compared with the thermal correlators.

\subsection{Evaluation of the reconstructed correlator}
The `reconstructed' Euclidean correlator is by definition
what the Euclidean correlator would be if the spectral 
function remained unchanged with respect to the 
\emph{in vacuo} spectral function.
We want to evaluate this reconstructed Euclidean correlator
(\eq\ref{eq:Grec1-main}) in order to compare it with, 
and eventually subtract it from the thermal correlator.
This evaluation is simplified by the fact that 
the $\theta$ correlator falls off very fast 
at zero temperature.
A first approximation is thus provided by 
(see \eq\ref{eq:Grec2})
\be
G_E^{\rm rec}(t,T;0)\approx G_E(t,T=0) + G_E(\beta-t,T=0)\,.
\la{eq:Grec-est}
\ee
The state of smallest energy contributing to the 
zero-temperature correlator is the lightest glueball.
Its mass and the matrix element 
$F_S\equiv \<{\rm vac}| {\cal O}_\theta|{\rm glueball}\>$
have been calculated on the lattice
(in this definition the norm of the states are set to unity).
When $\beta m$ is large, we can therefore 
estimate the leading correction to \eq(\ref{eq:Grec-est}),
\be
e^{-m\beta}\cdot \frac{F_S^2\cosh M(\beta/2-t)}{\sinh \beta M/2}\,.
\la{eq:Grec-M}
\ee
As examples we give the respective size of contributions
(\ref{eq:Grec-est}) and (\ref{eq:Grec-M})
to the reconstructed correlator (in units of $T^5$)
at a few temperatures of interest in the deconfined phase,
\ba
T=1.02T_c:&\qquad&  22.10(71)   \qquad  0.081(11)
\\
T=1.24T_c:&\qquad& 15.36(39) \qquad 0.117(16)  
\\
T=1.65T_c:&\qquad & 9.47(43)  \qquad 0.132(23)\,.  
\ea
As these examples show, (\ref{eq:Grec-M}) is a 
small correction to (\ref{eq:Grec-est}), in fact smaller
than the latter's statistical error. In subsequent figures we will 
therefore display (\ref{eq:Grec-est}) as a simple estimate
of the reconstructed correlator. We note that expression 
(\ref{eq:Grec-est}) is a strict lower bound on the 
reconstructed correlator.

\subsection{Large-frequency contributions to spectral sums}
For a large value of $\omega_2$, \eq(\ref{eq:OPE}) 
implies to leading order,
\be
\frac{2}{e+p}
\int_{\omega_2}^\infty \frac{d\omega}{\omega}\,\Delta\rho_\star(\omega,{\bf 0},T)
= - \left(\frac{11N\alpha_s(\omega_2)}{6\pi}\right)\,\frac{e-3p}{e+p}
  -  \frac{3}{4} \left(\frac{11N\alpha_s(\omega_2)}{6\pi}\right)^2 \,.
\la{eq:w.gt.w2}
\ee
For instance, if we choose $\omega_2=12/r_0\approx 4.8$GeV, 
then  $\alpha_s(\omega_2) \simeq 0.14$~\cite{Luscher:1993gh}, 
and 
the second term on the RHS of \eq(\ref{eq:w.gt.w2})
amounts to (-0.046). The first term then amounts to 
\be
-0.11,~~ -0.053,~~ -0.025,~~ -0.011
\ee
at 1.24, 1.65, 2.22 and 3.22$T_c$ respectively.
This evaluation shows that the two terms on the RHS of 
\eq(\ref{eq:w.gt.w2}) are about equally important at $T=2T_c$
and that the second term becomes dominant for higher temperatures.

The contribution of the tail of the spectral function to the 
Euclidean correlator at $t=\beta/2$ is given in leading approximation by
\be
\int_{\omega_2}^\infty \frac{d\omega}{T}\, 
\frac{\Delta\rho_\star(\omega,T)}{\sinh \omega/2T}
\simeq 4\Delta\rho_\star(\omega_2,T) \, e^{-\omega_2/2T}.
\ee
At $T=2T_c$, the exponential suppression is quite strong,
$e^{-\omega_2/2T} \approx 0.018$ and $\Delta\rho_\star(\omega_2)$ itself
is of order (-0.02). This makes the large-frequency contribution
to the Euclidean correlator negligible, and in any case much 
smaller than its current statistical error.

\subsection{The temperature dependence of the bulk spectral function}

\fig(\ref{fig:nearTc}) compares the thermal Euclidean 
correlators $C_\theta(t,T)$ and $C_\star(t,T)$
with the reconstructed correlator 
$C^{\rm rec}_\star(t,T;T_0)$.
We see again that the difference between 
$C_\theta$ and $C_\star$ is small, except close to $T_c$.
However, the difference between the 
reconstructed correlator and $C_\star$ is small too.
Therefore the $t$-independent shift between 
$C_{\star}$ and $C_\theta$ is nonetheless crucial to 
correctly evaluate 
$C_\star(t,T)- C^{\rm rec}_\star(t,T)$.
From \fig(\ref{fig:nearTc}), we learn in particular that 
the central value of this difference 
is positive at $t=\beta/2$. While on a finer lattice
one could reliably exploit more data points, here we 
will restrict ourselves to this middle-point, which carries
the most information on the low-frequency part of 
$\Delta\rho_\star$. Numerically we find,
in terms of the subtracted spectral function
(see \eq(\ref{eq:ClatRho}) and \eq(\ref{eq:Grec2})),
\ba
X(T) &\equiv& \frac{T^4}{e+p}\,
[C_\star(t,T)-C^{\rm rec}_\star(t,T) ]_{t=\beta/2}\,,
\\
X(T) &=& \frac{1}{e+p}\int_0^\infty \frac{d\omega}{T}\,
\frac{\Delta\rho_\star(\omega,T)}{\sinh \frac{\omega}{2T}}
= \left\{ \begin{array}{l@{\qquad}l}
0.7 \pm 3.1 & 1.02T_c\\
0.26 \pm 0.13   & 1.24 T_c \\
0.22\pm 0.10   & 1.65T_c.
\end{array}\right.
\la{eq:Cmid}
\ea
On the other hand, the bulk sum rule (\ref{eq:bsr1}) 
tells us that 
\be
Y(T) \equiv \frac{2}{e+p}\int_0^\infty \frac{d\omega}{\omega}
\,\Delta\rho_\star(\omega,T)  =
\left\{ \begin{array}{l@{\qquad}l}
-0.928\pm 0.038 & 1.02T_c\\
-1.204\pm 0.086 & 1.24T_c\\
-0.578\pm 0.099 & 1.65T_c\,.
\end{array}
\right.
\la{eq:bsr}
\ee
These two equations suggest an enhancement 
of the thermal spectral weight at low frequencies
and a suppression at the higher frequencies.
Indeed the higher frequencies contribute much more
to the bulk sum rule than to \eq(\ref{eq:Cmid}).

As a technical remark we note that since 
a large cancellation takes place when subtracting
the reconstructed correlator, the quantity $X(T)$
is susceptible to a large discretization error 
in addition to the quoted statistical error.
In obtaining the results (\ref{eq:Cmid}),
we have used treelevel-improvement~\cite{Meyer:2009vj} 
separately
for the finite-temperature and the zero-temperature
correlators before calculating $X(T)$.
In~\cite{Meyer:2007dy} it was shown that 
after treelevel-improvement 
the discretization errors affecting 
$C_\theta(t=\beta/2,T)$ are mild.
Also, we have ignored possible finite-volume effects,
which could potentially induce a large relative error in $X(T)$.
New data on finer and physically larger lattices 
will be necessary to obtain $X(T)$ with completely
controlled systematic errors.

\subsection{Thermal rearrangement of the spectral weight}
We have collected a certain amount 
of information on the thermal spectral function
relative to its zero-temperature counterpart.
Firstly, we know the asymptotic large-frequency behavior, 
thanks to the operator-product expansion.
Secondly, two integrals over all frequencies 
of the subtracted spectral function are known.
We also know that the difference $\Delta\rho_\star$ is positive 
for $\omega$ below the mass gap $m$, 
since the zero-temperature spectral function
vanishes in that region, and we know the position and strength
of the glueball peak.

When determining properties of the spectral function,
it is technically 
advantageous to separate the exact delta functions
it contains from the contributions which are smooth 
in the infinite volume limit.
Identifying the delta function at the origin 
in the bulk spectral function was an important step
in this direction. 
Since we are interested in low-energy properties,
it is also advantageous to work with a subtracted
spectral function, such that the difference goes 
to zero at high frequencies. This is achieved 
by subtracting the zero-temperature spectral function.
However, doing so one introduces, with negative weight,
delta functions that correspond to the stable 
one-particle states at zero-temperature~\cite{Meyer:2008dq}. 
In the pure SU($N$) gauge theory, these are the 
glueballs. The position and weight of these
delta functions can in principle be determined.
They have been determined for the lightest 
glueball~\cite{Meyer:2008tr}, 
but only rough estimates are known for the second 
(and last) stable scalar glueball.
Therefore we will not attempt to systematically isolate
the stable-glueball contributions in the spectral integrals
 $X(T)$ and $Y(T)$,
but it is instructive to estimate the negative contribution 
of the lightest scalar glueball, 
\ba
-\delta_{G}X(T) = \frac{F_S^2}{T(e+p)}\, 
\frac{1}{\sinh m/2T}
&=&\left\{\begin{array}{l@{\qquad}l}
1.6(2) &   1.24T_c\\
0.55(10) &   1.65T_c.
\end{array}\right.
\\
-\delta_{G}Y(T) =
\frac{2 F_S^2}{m(e+p)} 
&=&\left\{\begin{array}{l@{\qquad}l}
3.0(4) &   1.24T_c\\
0.81(14) &   1.65T_c.
\end{array}\right.
\ea
These contributions are quite 
large compared to $X(T)$ and $Y(T)$ themselves.
We have seen that $Y(T)$ is negative.
Based on the OPE, only a small fraction of its magnitude
can be explained by a depletion of the spectral density for
$\omega>\omega_2=12/r_0\approx4.8$GeV.
Therefore the bulk of the contribution to $Y(T)$ must 
come from the region $m\leq \omega \leq \omega_2$.
\emph{The depletion of the spectral density in the region
of those glueballs that are stable or lie 
slightly above the two-particle threshold
must therefore be the explanation for the 
sign and magnitude of $Y(T)$.} Whether the glueballs 
are completely `melted' in the deconfined phase,
in the sense that the spectral function shows 
no narrow peaks in this region, cannot be answered
on the basis of the available information.

\FIGURE[t]{
\centerline{\includegraphics[width=6.0 cm,angle=-90]{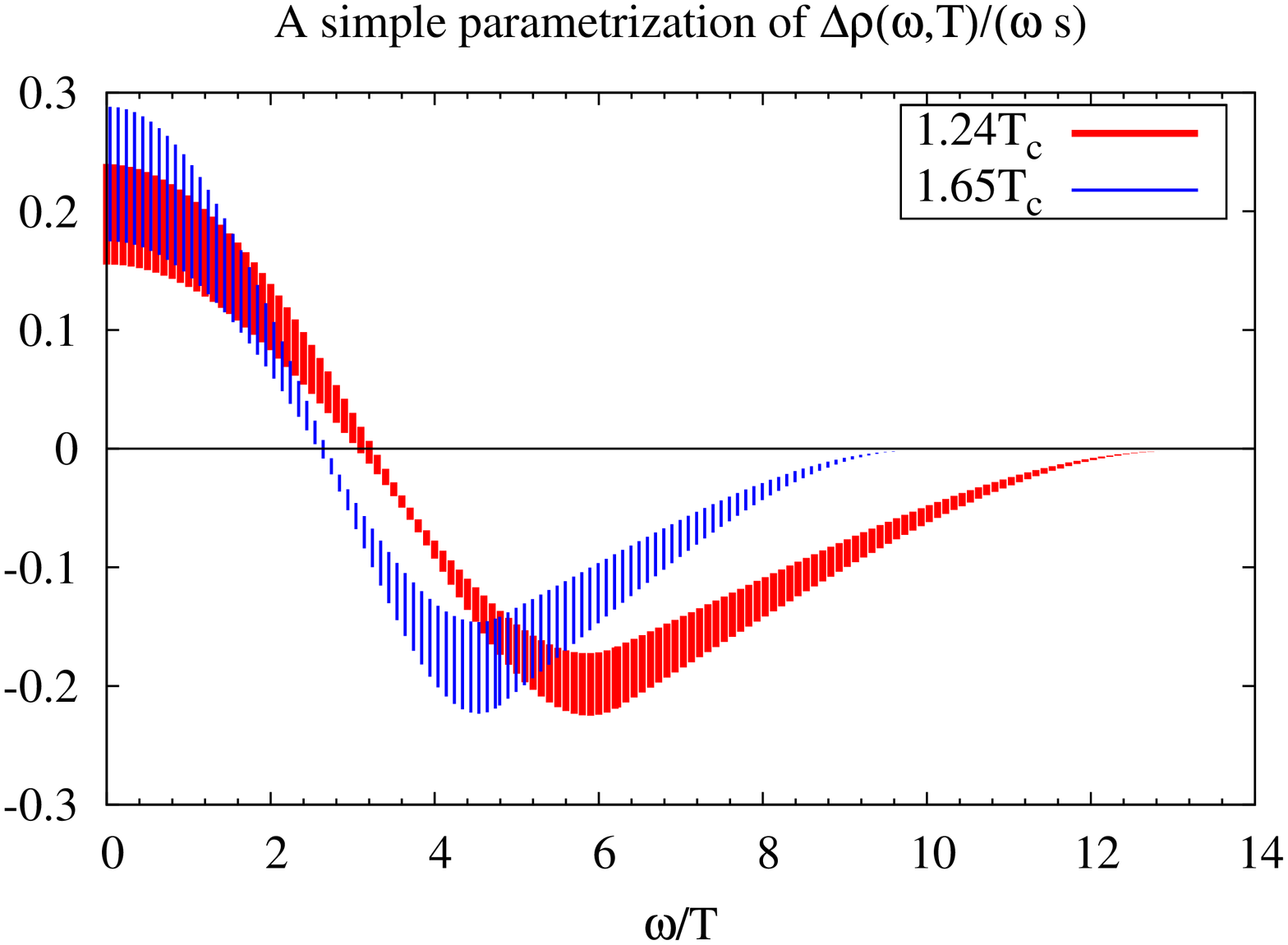}}
\caption{
A smooth parametrization of the thermal part 
of the subtracted spectral function,
$\Delta\rho_\star(\omega,T)/(\omega\cdot s)$, 
compatible with the bulk sum rule
and the lattice correlator at $t=\beta/2$. 
Recall that the ratio $\zeta/s$ is given by 
$\frac{\pi}{9}\times$ the intercept of this function.}
\label{fig:recon}
}
 \FIGURE[t]{
\vspace{-0.2cm}

 \centerline{\includegraphics[width=6.8 cm,angle=-90]{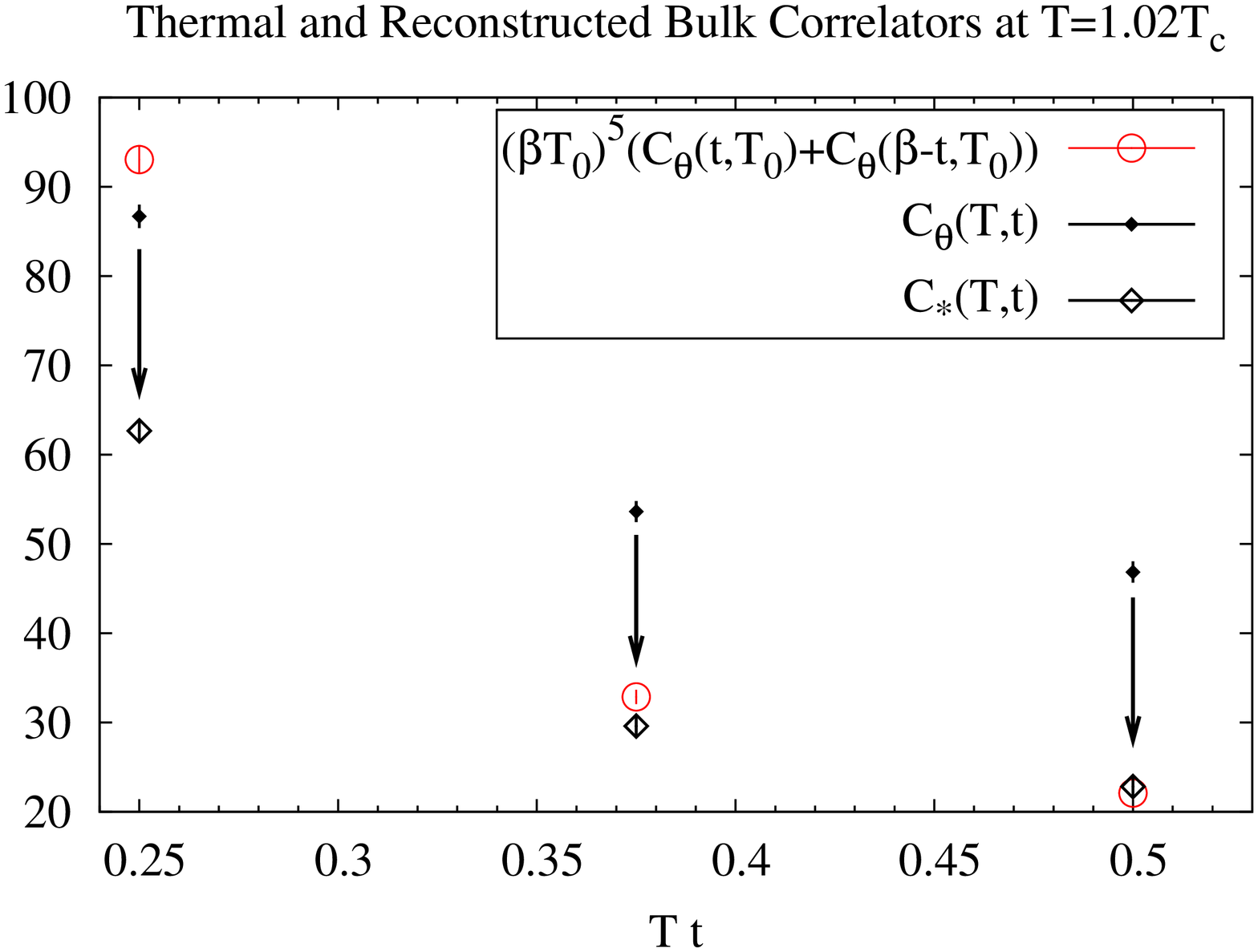}}
\vspace{-0.2cm}

 \centerline{\includegraphics[width=6.8 cm,angle=-90]{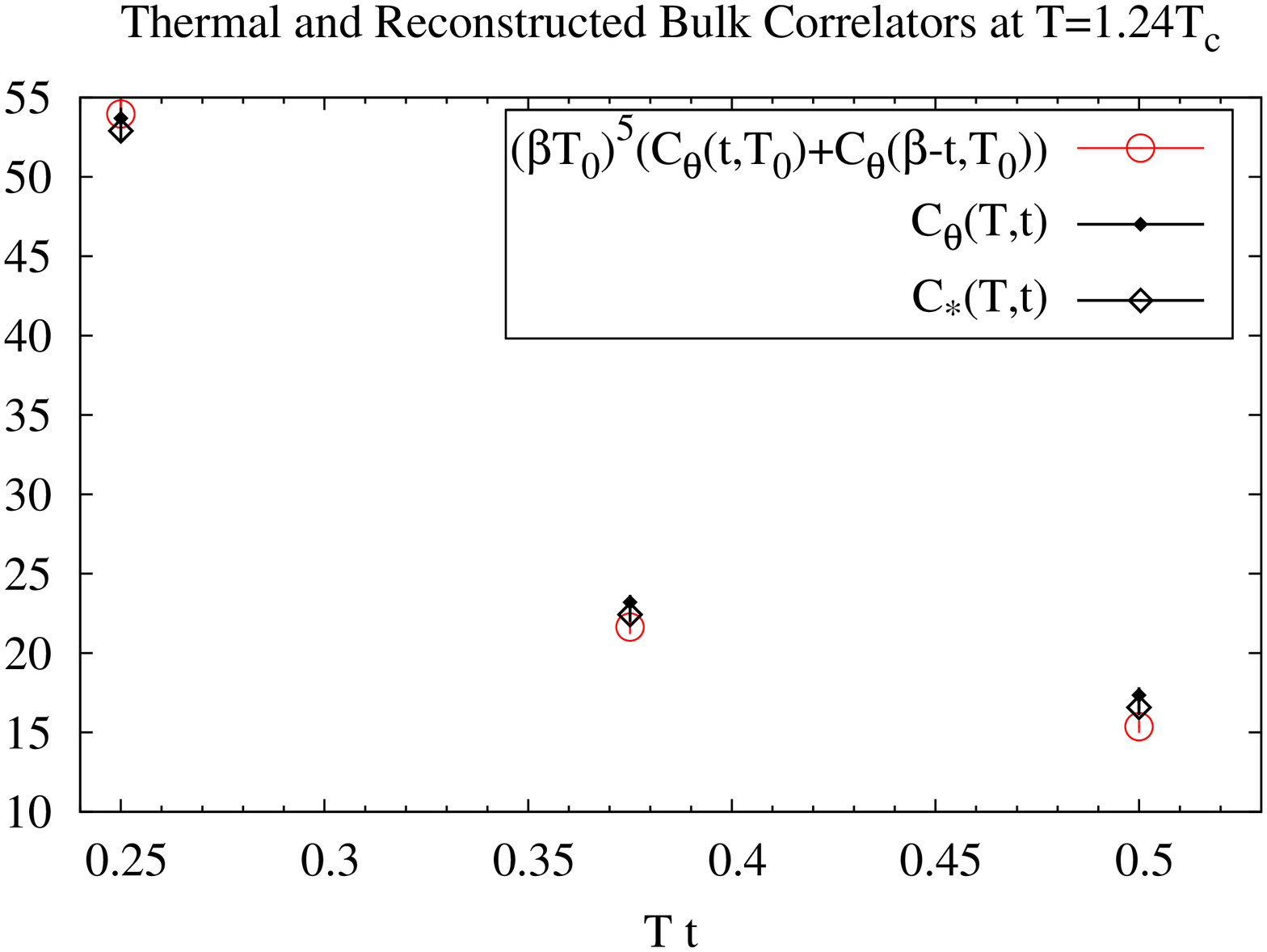}}
\vspace{-0.2cm}

 \centerline{\includegraphics[width=6.8 cm,angle=-90]{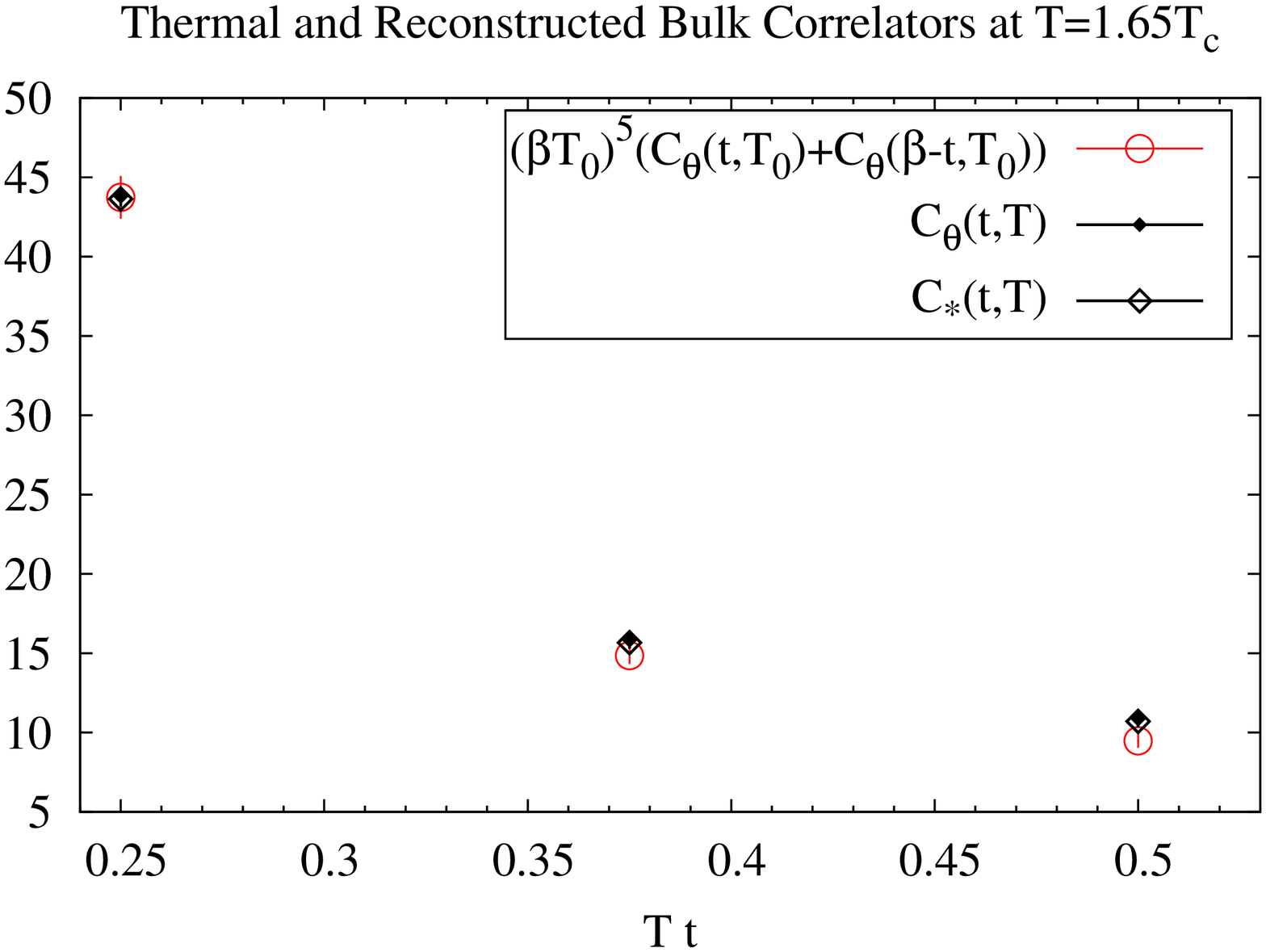}}
 \caption{
Bulk-channel correlators at three temperatures.
Here $\Nt=8$ and treelevel-improvement has been applied.
The open circles correspond to what the Euclidean correlator would be 
if the spectral density remained unchanged from a much lower temperature,
$T_0\simeq T_c/2$.
}
 \label{fig:nearTc}
 }

While no detailed properties of $\Delta\rho_\star$ can be 
determined based on the limited information at hand,
we want to show somewhat more quantitatively
how \eq(\ref{eq:Cmid}) and (\ref{eq:bsr}) 
dictate the rearrangement of the spectral weight.
For instance, in an interval of frequencies, one may ask whether
$\int_{\omega_1}^{\omega_2} \frac{d\omega}{\omega}\Delta\rho_\star$ 
is positive or negative.
In order to address this question in a systematic way,
we apply the following procedure (it is of course not unique).
We select a binning in the frequency variable $\omega$.
We choose $[0,\omega_1]$ and $[\omega_1,\omega_2]$.
The upper end of the second bin, $\omega_2$, is chosen 
large enough for the OPE to be reliable.
Since $\rho_\star(\omega,T=0)$ vanishes for $\omega<m$, 
it is of physical interest to choose $\omega_1$ on the order
of $m$.

Second, since we expect the (infinite-volume)
spectral function to be smooth at least at high-frequencies,
we parametrize $\Delta\rho_\star$ by a second-order polynomial 
in $\omega^2$ in each bin. We require continuity 
and differentiability at the boundaries between the bins.
For $\omega>\omega_2$, we use the expression provided by the OPE.
This then leaves 6 coefficients to be determined. 
Four equations are provided by the matching conditions at the 
edges of the bins, and two are provided by the known frequency integrals
of $\Delta\rho_\star$, \eq(\ref{eq:Cmid}) and (\ref{eq:bsr}). 
Determining this simple parametrization of the 
spectral function thus amounts, in the present case,
to inverting a $6\times6$ matrix.

The result of this procedure is show in \fig(\ref{fig:recon}).
The bands indicate the statistical error obtained by 
standard error propagation.
These curves show qualitatively what was expected from 
the remarks below \eq(\ref{eq:bsr}),
namely the appearance of a positive spectral weight
for $\omega<m$, and a depletion relative to the vacuum
spectral density for $\omega>m$. Incidentally,
this is qualitatively consistent with the prediction
of perturbation theory, where the O($\alpha_s^2$) 
low-frequency and O($\alpha_s^2$) high-frequency contributions 
cancel eachother, resulting in a spectral sum
$\int d\omega\Delta\rho_\star/\omega$ that is O($\alpha_s^3$)
\cite{Romatschke:2009ng,CaronHuot:2009ns}.

We note that the values of the intercept in \fig(\ref{fig:recon})
at $1.24T_c$ and $1.65T_c$,
although strongly dependent on the fact that we have 
smoothened the spectral function, are in good agreement 
with our previous estimates for $\zeta/s$~\cite{Meyer:2007dy},
where the unsubtracted spectral function $\rho_\theta$ 
was directly parametrized. This may well be because 
in both cases we made a smooth ansatz for the spectral
function, but in view of the large cancellation that occurs
in $\Delta\rho_\star$, the consistency is not totally trivial.

\section{Conclusion}

We have collected information, obtained 
by analytic methods, on the 
spectral density of the bulk channel
in SU($N$) gauge theory.
We have reviewed
the precise connection between the Euclidean 
and the retarded correlator in frequency space, 
illustrated in \fig(\ref{fig:sketch}), recalculated
the OPE of the bulk channel correlator, \eq(\ref{eq:OPE}),
and presented a new way, based on (\eq\ref{eq:Grec2}),
to compute the reconstructed 
correlator. The latter technique allows one to 
directly constrain the difference of thermal to 
vacuum spectral densities, $\Delta\rho$.

Numerically, we have found that in the deconfined phase 
of SU(3) gauge theory, a significant depletion of the 
spectral density must take place in the interval 
$m\lesssim\omega\lesssim 3m$, while we have presented
evidence that a non-vanishing
spectral density appears at frequencies below the mass gap,
obviously a finite-temperature effect. 
This qualitative conclusion comes from combining
the bulk sum rule with numerical lattice correlators.
With numerical data of higher accuracy, it would be possible
to quantify more precisely how much spectral weight
is clustered around the origin with
frequencies $\omega\lesssim T$.

Unfortunately, the behavior of the bulk spectral function
at low frequencies near the deconfining temperature
remains largely undetermined. The Euclidean correlator 
$C(t,T\approx T_c)$
clearly exhibits an enhancement compared to higher temperatures,
even after subtraction of the $t$-independent 
contribution, see \eq(\ref{eq:Cstar}), which is quite 
large just above  the transition. However
this enhancement can be explained to a large extent
by the properties of the vacuum, 
as the comparison with the reconstructed correlator shows. 
At the moment we can state that a depletion of the 
spectral density has to take place for frequencies 
above the mass gap, in view of the bulk sum rule, but 
are unable at present to give a useful estimate of 
the spectral weight below the mass gap.
This more careful reanalysis of the bulk channel has 
taught us that the effects of a potentially large bulk
viscosity near $T_c$ are more subtle to detect 
in spectral integrals than previous work 
suggested~\cite{Kharzeev:2007wb,Meyer:2007dy}.
More accurate lattice data, in particular for the 
speed of sound near the phase transition, would allow one
to determine the cumulated spectral weight below the mass gap.
We believe that this goal can be achieved in the foreseeable future.

\acknowledgments{
I thank K.~Rajagopal and U.~Wiedemann 
for interesting discussions on the bulk channel and 
G.D.~Moore, D.T.~Son, S.S.~Gubser,
S.S.~Pufu and F.D.~Rocha for a correspondence on the 
bulk spectral density.
The simulations were carried out on a Blue Gene L rack 
and on desktops of the Laboratory for Nuclear Science, 
Massachusetts Institute of Technology.
This work was supported in part by
funds provided by the U.S. Department of Energy 
under cooperative research agreement DE-FG02-94ER40818.
}
\appendix

\input{apdx.tex}

\bibliographystyle{JHEP}
\bibliography{../../../../BIBLIO/viscobib.bib}

\end{document}

%% file: apdx.tex
\section{Spectral representation of finite-temperature correlators}
In this appendix we give the spectral representation 
of various correlators. We use this representation
to obtain \eq(\ref{eq:GR+GE-main}).

\subsection{Real-time correlators}
Inserting two complete sets of energy eigenstates
in the expression (\ref{eq:defG_R-main}) for $G_R(t)$, 
one obtains
\be
G_R(t) = \frac{i\theta(t)}{Z}
\sum_{m,n} |O_{nm}|^2 e^{-\beta E_n}(e^{iE_{nm}t}-e^{-iE_{nm}t})\,,
\ee
where we employ the notation
\be
E_{nm}=E_n-E_m\,,\qquad O_{nm} = \< n | \hat\O(t=0) | m\>\,.
\ee
In the same representation, the Fourier transform 
(\ref{eq:defG_Rtilde-main}) reads, for $\im \omega>0$,
\ba
\tilde G_R(\omega) &=&
\frac{1}{Z}\sum_{m,n}  |O_{nm}|^2 e^{-\beta E_n}
\left(\frac{1}{\omega-E_{nm}} - 
      \frac{1}{\omega+E_{nm}} \right)
\la{eq:G_R1}
\\
&=& 
\frac{1}{Z}\sum_{m,n}  |O_{nm}|^2 e^{-\beta E_n}
\frac{2E_{nm}}{\omega^2-E_{nm}^2}\,.
\la{eq:G_R2}
\ea
In particular, for later reference, 
\ba
\tilde G_R(i\omega_{\rm I})& =  &
-\frac{1}{Z}
\sum_{n,m}  |O_{nm}|^2 e^{-\beta E_n} 
\frac{2E_{nm}}{\omega_{\rm I}^2+E_{nm}^2}\,
\\
&=&
\frac{\beta}{Z}\sum_{n,m}  |O_{nm}|^2  e^{-\beta(E_n+E_m)/2}
\frac{E_{nm}^2}{\omega_{\rm I}^2+E_{nm}^2}
\frac{\sinh(\beta E_{nm}/2)}{\beta E_{nm}/2}
\quad (\omega_{\rm I}>0).
\ea
\subsection{Euclidean correlators}
The spectral representation of the Euclidean correlator
(\ref{eq:GE}) reads
\be
G_E(t) = \frac{1}{Z} \sum_{n,m} |O_{nm}|^2 
        e^{-\beta E_n} e^{E_{nm} t} \,,
\la{eq:G_E1}
\ee
and after some algebraic manipulations,
one also finds for the Fourier coefficients (\ref{eq:GEell})
\be
G_E^{(\ell)} = 
\frac{\beta}{Z}\sum_{m,n} 
|O_{nm}|^2 e^{-\beta(E_n+E_m)/2}
\frac{E_{nm}^2}{\omega_\ell^2+E_{nm}^2}
\frac{\sinh(\beta E_{nm}/2)}{\beta E_{nm}/2}
\la{eq:G_El}
\ee
Thus in particular we see that \eq(\ref{eq:l.neq.0-main}) holds.

The relation of the Matsubara zero-mode $\ell=0$ with 
the zero-frequency retarded correlator requires special care. 
Using the spectral representation, one finds
\ba
G_E^{(0)} - G_R(i\omega_{\rm I})
&=&
\frac{\beta}{Z}\sum_{m,n} |O_{nm}|^2 e^{-\beta (E_n+E_m)/2}
\frac{\sinh(\beta E_{nm}/2)}{\beta E_{nm}/2}
\frac{\omega_{\rm I}^2}{\omega_{\rm I}^2+E_{nm}^2}\,.
\la{eq:l.eq.0a}
\ea
In the next section, we shall see that the right-hand side can
be expressed in terms of the spectral function.

\subsection{The spectral function, and its relation to 
Euclidean correlators}
Taking the imaginary part of the retarded correlator,
\eq(\ref{eq:G_R1}), one obtains
\be
\im\tilde G_R(\omega) =
\frac{1}{Z}\sum_{n,m} |O_{nm}|^2 e^{-\beta E_n}
\left(\frac{\omega_{\rm I}}{(\omega_{\rm R}+E_{nm})^2+\omega_{\rm I}^2}
 -\frac{\omega_{\rm I}}{(\omega_{\rm R}-E_{nm})^2+\omega_{\rm I}^2}\right)
\la{eq:ImGR}
\ee
For a complex frequency just above the real axis, 
$\omega_{\rm I}=\epsilon$, we obtain
\be
\im\tilde G_R(\omega+ i\epsilon) =
\frac{\pi}{Z}\sum_{n,m} |O_{nm}|^2 e^{-\beta E_n}
[\delta(\omega+E_{nm}) - \delta(\omega-E_{nm})]\,,
~~ \omega\in {\bf R}.
\la{eq:imGR}
\ee
We have used one of the standard representations
of the delta function,
\be
\delta(\omega)=\frac{1}{\pi} \frac{\epsilon}{\omega^2+\epsilon^2}\,.
\la{eq:delta}
\ee
Symmetrizing the sum in \eq(\ref{eq:imGR}) 
with respect to $m$ and $n$, one easily reaches the form
($\omega\in{\bf R}$)
\be
\rho(\omega) = \frac{1}{\pi}\lim_{\epsilon\to 0}
{\rm Im} \tilde G_R(\omega+i\epsilon)
= \frac{2}{Z}\, \sin\frac{\beta\omega}{2}
\sum_{m,n}  |O_{nm}|^2 e^{-\beta (E_n+E_m)/2}
\delta(\omega-E_{nm})\,.
\la{eq:rho}
\ee
We thus observe that the limit of
$\tilde G_R$ for $\omega$ approaching the real axis 
is in general a distribution.
\begin{enumerate}
\item
It is easy to verify, using \eq(\ref{eq:G_E1}) and (\ref{eq:rho}),
that the Euclidean correlator can be obtained from $\rho$ via
\eq(\ref{eq:ClatRho}).
\item
Substituting (\ref{eq:rho}) into the right-hand side of 
\eq(\ref{eq:GR+GE-main}),
one recovers \eq(\ref{eq:l.eq.0a}), thus proving 
\eq(\ref{eq:GR+GE-main}).
\item A simple calculation based on expression (\ref{eq:rho})
formally leads to \eq(\ref{eq:KK-main}).
\end{enumerate}